\RequirePackage{lineno}
\documentclass[reprint,aps]{revtex4-1}
\usepackage[%
  colorlinks=true,
  urlcolor=blue,
  linkcolor=blue,
  citecolor=blue
]{hyperref}
\usepackage{amsthm}
\usepackage{subfigure}
\usepackage{graphicx}
\usepackage{epsfig}
\usepackage{epstopdf}
\usepackage{color}
\usepackage{caption}
\usepackage{amsmath}
\usepackage{amssymb}
\usepackage{hyperref}
\usepackage[shortlabels]{enumitem}
\usepackage{color}
\usepackage{soul}

\def \but {\boldsymbol{\widetilde{u}}}

\def \bcdot {\boldsymbol{\cdot}}

\def \bex {\boldsymbol{e}_x}

\def \bu {\boldsymbol{u}}

\def \pa {\partial}
 
\def \xt {\widetilde{x}}

\def \wt {\widetilde}

\def \tt {\wt{t}}

\def \s0 {\s_0}
\def \s {\wt{\sigma}}

\def \bseq {\begin{subequations}}
\def \eseq {\end{subequations}}

\def \lap {\nabla^2} 
\def \lapt {\widetilde{\nabla}^2} 
\def \bseq {\begin{subequation}}
\def \eseq {\end{subequation}}
\def \beq {\begin{equation}}
\def \eeq {\end{equation}}
\def \beqn {\begin{eqnarray}}
\def \eeqn {\end{eqnarray}}
\def \bi {\begin{itemize}}
\def \ei {\end{itemize}}
\def \be {\begin{enumerate}}
\def \ee {\end{enumerate}}
\def \bfig {\begin{figure}}
\def \efig {\end{figure}}
\def \ba {\begin{align}}
\def \ea {\end{align}}

\def \bseq {\begin{subequations}}
\def \eseq {\end{subequations}}
\def \k {\kappa}

\def \bnabla {\boldsymbol{\nabla}}

\def \cmt {\widetilde{c}_m}
\def \cst {\widetilde{c}_s}
\def \mt {\widetilde{\mu}}
\def \tt {\widetilde{t}}
\def \yt {\widetilde{y}}
\def \pt {\widetilde{p}}
\def \mut {\widetilde{\mu}}

\newcommand{\bra}[1]{\left( #1 \right)}
\newcommand{\sqbra}[1]{\left[ #1 \right]}

\newcommand{\grad}[1]{\boldsymbol{\nabla} #1}
\newcommand{\divg}[1]{\bnabla \bcdot #1}

\newcommand{\gradt}[1]{\boldsymbol{\widetilde{\nabla}} #1}
\newcommand{\divgt}[1]{\widetilde{\bnabla} \bcdot #1}

\begin{document}


\title{Influence of Langmuir adsorption and viscous fingering on transport of finite size samples in porous media}

\author{C. Rana\textsuperscript{1,2}, S. Pramanik\textsuperscript{1,3}, M. Martin\textsuperscript{4}, A. De Wit\textsuperscript{2} and M. Mishra\textsuperscript{1,5}} 
\affiliation{{\textsuperscript{1}Department of Mathematics, Indian Institute of Technology Ropar, Rupnagar-140001, Punjab, India}} \affiliation{ {\textsuperscript{2}Universit\'e Libre de Bruxelles (ULB), Nonlinear Physical Chemistry Unit, 1050 Brussels, Belgium}} \affiliation{{\textsuperscript{3}Nordita, Royal Institute of Technology and Stockholm University, SE 106 91 Stockholm, Sweden}}\affiliation{{\textsuperscript{4}PMMH, ESPCI Paris, CNRS, PSL (Paris Sciences et Lettres) University, Sorbonne-Universit\'e, Universit\'e Paris-Diderot,  Campus Jussieu, Barre Cassan A Case 18 7 Quai Saint-Bernard 75252 Paris Cedex 05, France}} \affiliation {{\textsuperscript{5}Department of Chemical Engineering, Indian Institute of Technology Ropar, Rupnagar-140001, Punjab, India}  }

\date{\today}

\begin{abstract}
We examine the transport in a homogeneous porous medium of a finite slice of a solute which adsorbs on the porous matrix following a Langmuir adsorption isotherm and can influence the dynamic viscosity of the solution. In the absence of any viscosity variation, the Langmuir adsorption induces the formation of a shock layer wave at the frontal interface and of a rarefaction wave at the rear interface of the sample. For a finite width sample, these waves interact after a given time that varies nonlinearly with the adsorption properties to give a triangle-like concentration profile in which the mixing efficiency of the solute is larger in comparison to the linear or no-adsorption cases. In the presence of a viscosity contrast such that a less viscous carrier fluid displaces the more viscous finite slice, viscous fingers are formed at the rear rarefaction interface. The fingers propagate through the finite sample to preempt the shock layer at the viscously stable front. In the reverse case i.e. when the shock layer front features viscous fingering, the fingers are unable to intrude through the rarefaction zone and the qualitative properties of the expanding rear wave are preserved. A non-monotonic dependence with respect to the Langmuir adsorption parameter $b$ is observed in the onset time of interaction between the nonlinear waves and viscous fingering. The coupled effect of viscous fingering at the rear interface and of Langmuir adsorption provides a powerful mechanism to enhance the mixing efficiency of the adsorbed solute. 

\end{abstract}

\maketitle 
\section{Introduction}
Waves are ubiquitous in a wide variety of physical and chemical systems ranging from geophysical fluid dynamics \cite{Miles1957, Vanneste2013}, aerodynamics \cite{Crighton1979}, the motion of glaciers and traffic flow \cite{Whitham1999}, separation science \cite{Guiochon2006}, quantum mechanics \cite{Steeb1998}, and astrophysics \cite{McKee1980}, among others. Although nonlinear waves are quite complex, some of them are analytically tractable. \emph{Shock layer} (SL) and \emph{rarefaction} (RF) waves are two such classes of nonlinear waves which have been studied extensively using analytic/semi-analytic methods. A SL is a nonlinear wave with a steep but continuous profile while a RF wave is a nonsharpening wave with a highly diffused profile \cite{Helfferich1993}. Interactions of these nonlinear waves lead to many interesting nonlinear dynamics and pattern formation. A \emph{triangular wave} (or \emph{N wave}) forms when a SL interacts with a RF wave \cite{Whitham1999}. From the theoretical perspective, the study of triangular waves is of fundamental interest in many physico-chemical systems where the equations of motion possess both shock layer and rarefaction solutions \cite{Evans1962, Ruthven1984}. Here, we investigate analytically and numerically the interaction between SL and RF waves in a porous matrix, e.g., a soil or a chemical reactor, when nonlinear adsorption-desorption processes impact the transport of given solutes \cite{Guiochon2006,Weber1991}. The adsorption isotherm describes the partitioning of solutes between the solvent (mobile phase) and the matrix/adsorbent (stationary phase). For a finite slice of a solute undergoing a fluid-solid Langmuir adsorption isotherm \cite{Langmuir1916}, the stationary phase concentration $\cst$ varies with the concentration in the mobile phase $\cmt$ as : 
\begin{equation}
\label{eq:Langmuir_isotherm}
\cst = \frac{K\cmt}{(1 + \bar{b}\cmt)}, 
\end{equation}
where $K$ is the equilibrium constant, while $\bar{b} = K/c_{sat}$ represents the rate at which $\cst$ saturates to $c_{\rm sat}$ when $\cmt$ increases. A Langmuir isotherm induces $\cmt$-dependent transport coefficients of the solute, which can lead to the sharpening or spreading of the solute concentration front, depending on the initial profile of $\cmt$. The transport equation of $\cmt$ satisfies then wave-like solutions \cite[and refs. therein]{DeVault1943, Helfferich1993, Rana2017a}. If $\cmt$ decreases in the direction of the wave motion, a SL wave forms \citep{Rana2017a} whereas, when $\cmt$ increases along the direction of the wave, a RF wave builds up. Thus for a finite solute slice, a Langmuir adsorption induces a SL (RF) formation at the frontal (rear) interface of the solute zone hence forming a triangle-like concentration profile \cite{Shirazi1988, Lin1995, Weber1991}.

Gradients of concentration across such waves can induce a hydrodynamic instability like viscous fingering (VF) for instance \cite{Homsy1987,Tan1988, DeWit2005}. VF develops when a less viscous fluid displaces a more viscous one and is of importance in contaminant transport, separation processes, enhanced oil recovery and carbon dioxide capture \cite{Boulding2004, Guiochon2006, Berkowitz2008, Sayari2010}. VF has also been used as a tool to enhance fluid mixing \cite{Jha2011}. For a linearily adsorbed solute undergoing VF, experimental \cite{dickson} and numerical \cite{Mishra2007} analysis reveal that the linear adsorption slows down the growth of instability. For a nonlinear Langmuir isotherm, the theoretical analyses of the displacement of a semi-infinite sample by a semi-infinite displacing liquid reveal an early onset of VF if a SL is formed \cite{Rana2017a} and a delayed onset of VF if a RF builds up \cite{Rana2015}. For a finite width sample, a recent experimental work on a system with Langmuir adsorption shows that an additional band broadening effect can be due to VF \cite{Enmark2015}, yet it is not known how the interaction between nonlinear waves (RF and SL) is affected by VF and whether these waves persist or impede after the interaction.

\begin{figure}
\centering
\includegraphics[trim = 0.2in 2.3in 0.4in 1in, clip, scale=0.35]{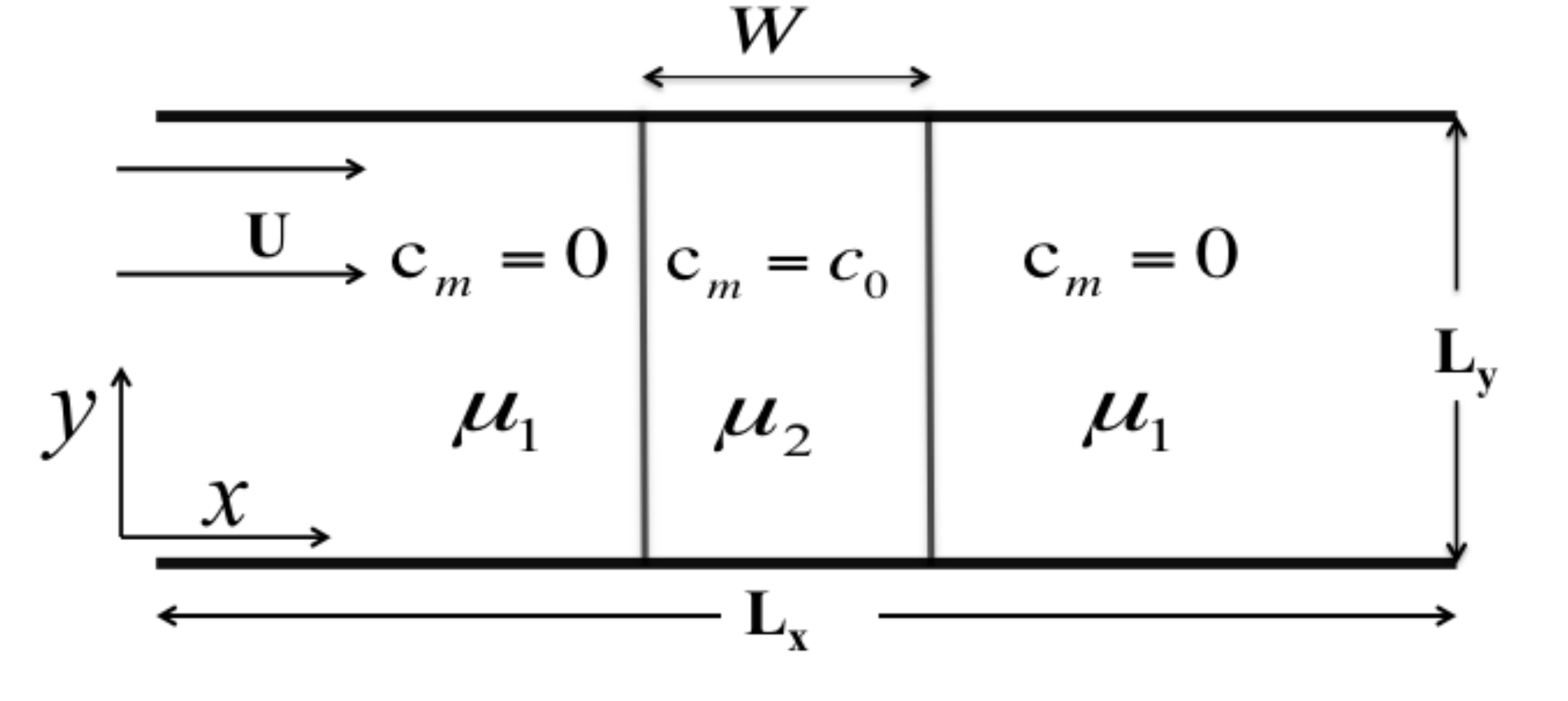}
\caption{Schematic of the displacement of a finite slice of miscible fluid in a two- dimensional homogeneous porous media.} 
\label{fig:R0_sample} 
\end{figure}

In this context, we analyse through mathematical analysis and numerical simulations the dynamics resulting from the interaction of RF and SL fronts during the displacement of a Langmuir adsorbed solute initially present in a finite width sample. In addition, we analyse the influence of viscous fingering on the fate of these nonlinear waves.

\section{Mathematical model}
We consider a rectangular porous medium of length $L_x$, width $L_y$, with a constant permeability $\kappa$. A fluid of viscosity $\mu_1$ displaces at a uniform velocity $U$ along the $x$-direction a sample of finite width $W$ of the same fluid containing a solute with initial mobile phase concentration $c_0$ and of viscosity $\mu_2$ (see Fig. \ref{fig:R0_sample}). The solute adsorbs on the porous matrix according to the Langmuir isotherm (\ref{eq:Langmuir_isotherm}). The governing equations for the solute transport, with a velocity field governed by Darcy's law, are:

\begin{eqnarray} \label{eq:1}
\divgt{ \but } &=& 0, \\ 
\gradt{p} &=& -\frac{\mt(\cmt)}{\k} \but, \label{eq:Darcy}\\
\label{eq:mass_conservation}
\frac{\partial \cmt}{\partial t}+F\frac{\partial \cst}{\partial t} +(\but  \cdot \gradt) \cmt &=& D\nabla ^2 \cmt,
\end{eqnarray}
where $\but=(u,v)$, $p$ is the pressure, $\mt(\cmt)$ is the dynamic viscosity of the fluid which depends on the mobile phase concentration $\cmt$, $\kappa$ is the permeability assumed here to be constant, $D$ is the dispersion coefficient of the solute in the solvent and $F=V_s/V_m$ is the phase ratio volume of the solute in the stationary and mobile phases. Substituting $\cst$ from Eq. (\ref{eq:Langmuir_isotherm}) into Eq. (\ref{eq:mass_conservation}), we get 
 \begin{equation} \label{eq:mass_conservation1}
 \frac{\pa }{\pa \tt} \sqbra{ \bra{ 1 + \frac{k}{1 + \bar{b} \cmt} } \cmt } + \but \bcdot \gradt{ \cmt } = D \lapt \cmt, 
 \end{equation}
where $k=FK$ is the retention parameter of the solute. The nonlinear dynamics of the solute concentration and the effect of viscous fingering can be analyzed by solving Eqs. \eqref{eq:1}-\eqref{eq:Darcy} and Eq. (\ref{eq:mass_conservation1}) subject to the following boundary and initial conditions, 
\begin{subequations}
\begin{align}
& \but = (U, 0), \quad \pa_{\xt} \cmt = 0, \qquad \mbox{as} \; \xt \to \pm \infty, \label{eq:BC1} \\ 
& \but(\xt, 0, \tt) = \but(\xt, L_y, \tt), \; \cmt(\xt, 0, \tt) = \cmt(\xt, L_y, \tt), \label{eq:BC2} \\ 
& \but = (U, 0) \label{eq:IC_u}, 
\end{align}
\end{subequations}
and an initial solute concentration in mobile phase $\cmt = c_0$ within the finite slice and $\cmt = 0$ outside it.

Using the following scalings 
\begin{subequations}
\begin{align}
& (x, y) = \frac{(\xt, \yt)}{D/U}, \quad t = \frac{\tt}{D/U^2}, \quad \bu = \frac{\but}{U}, \label{eq:scaling1} \\ 
& p = \frac{\pt}{\mu_1 D/\kappa}, \quad \mu = \frac{\mut}{\mu_1}, \quad c_m = \frac{\cmt}{c_0}, \label{eq:scaling2}
\end{align}
\end{subequations}
we obtain the following system of dimensionless equations, in a reference frame moving with the injection velocity: 
\begin{eqnarray} \label{eq:nondimgov}
& & \divg{\bu} = 0, \label{eq:nondim_continuity} \\ 
& & \grad{p} = -\mu(c_m)[\bu + \bex], \label{eq:nondim_Darcy} \\ 
& & \pa_{t} c_m + \sqbra{ \mathcal{D} (c_m) \bu - \mathcal{U}(c_m)\bex } \bcdot \grad{ c_m } = \mathcal{D}(c_m) \lap c_m, 
\nonumber\\ \label{eq:nondim_mass_conservation} 
\end{eqnarray}
where 
\begin{equation}
 \mathcal{U}(c_m) = \frac{k}{k + (1 + bc_m)^2}, \; \mathcal{D}(c_m) = \frac{(1 + b c_m)^2}{k + (1 + b c_m)^2}.  \label{eq:U}
\end{equation}
Here, $\bex$ is the unit vector along the $x$-axis and $b = \bar{b} c_0$ characterizes the Langmuir adsorption. For $b=0$, we recover the linear adsorption isotherm while $b \to \infty$ gives the no adsorption case.
The non-dimensional boundary and initial conditions in the moving reference frame are:
\begin{eqnarray}
& \bu = (0, 0), \; \partial_{{x}} c_m = 0 \; \mbox{at} \; x = 0, A \cdot {\rm Pe}, \label{eq:longitudinal_BC} \\ 
& \bu(x, 0, t) = \bu(x, {\rm Pe}, t), \; c_m(x, 0, t) = c_m(x, {\rm Pe}, t), \label{eq:transverse_BC} \\ 
& \bu(t = 0) = (0, 0), \label{eq:vel_IC} \\
& c_m = \left\{\begin{array}{ll}
1, & -l/2 \leq x \leq  l/2 \\
0, & \text{otherwise}
\end{array} \right.,  \label{eq:cm_IC}
\end{eqnarray}
where $\displaystyle l = \frac{W}{D/U}$ is the non-dimensional width of the finite slice of solute, $\displaystyle {\rm Pe} = \frac{UL_y}{D}$ is the P\'eclet number and $\displaystyle A = \frac{L_x}{L_y}$ is the aspect ratio of the system.

\begin{figure*}[!htbp]
\begin{tabular}{cc}
(a) & (b)\\
\includegraphics[scale=0.44]{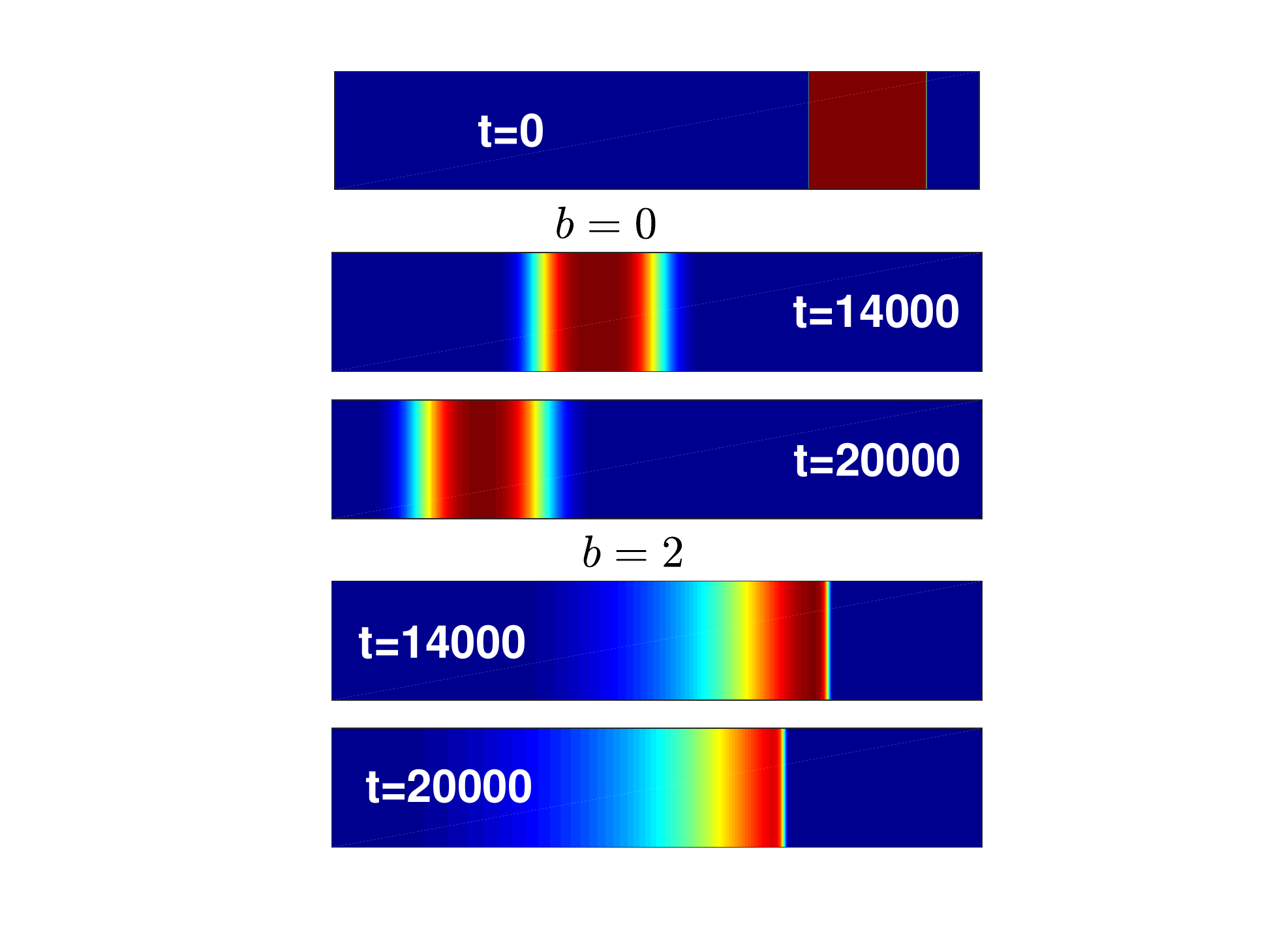} & \includegraphics[scale=0.4]{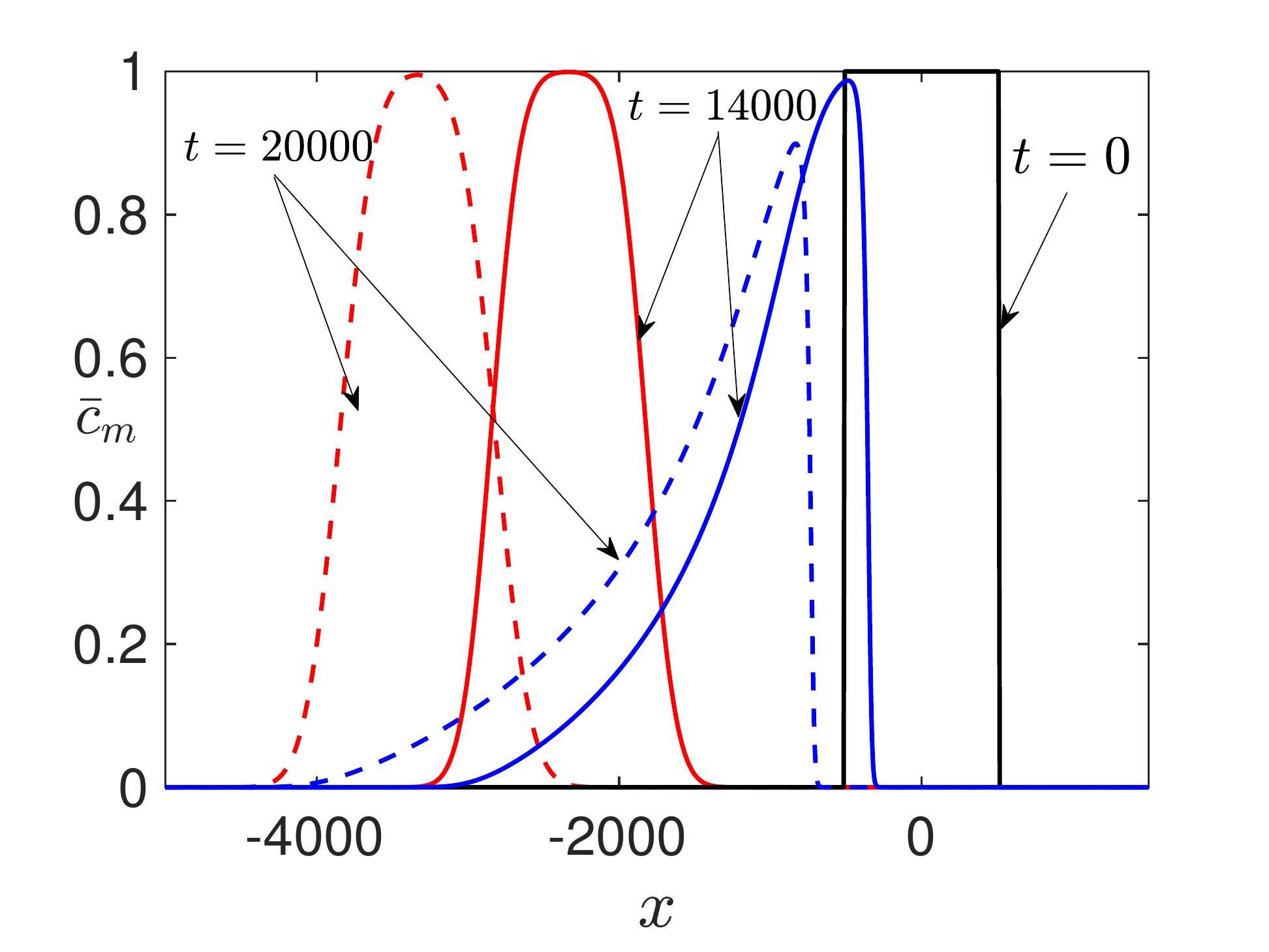}
\end{tabular}
\caption{(a) Concentration field $c_m(x,y,t)$ for $R=0,k=0.2$ with $b=0$ and $b=2$ at different times. (b) Corresponding transversely averaged concentration profiles $\bar{c_m}(x,t)$ for $R=0,k=0.2$ and $b=0$ (red), $b=2$ (blue).}
\label{fig:avg_b0b2_r0}
\end{figure*}  

The viscosity of the fluid is assumed to vary as
\begin{equation}
\mu(c_m) = \exp({R c_m}),
\end{equation}
 where $R = \ln(\mu_2/\mu_1)$ is the log-mobility ratio. If $R=0$, we have $\mu_2=\mu_1$ and the system does not exhibit any viscous fingering instability while, for $R>0$ the frontal interface and for $R<0$ the rear interface of the sample is viscously unstable \cite{DeWit2005, Mishra2008}. In order to analyse the propagation dynamics of the adsorbed solute, the stream-function vorticity form of equations (\ref{eq:nondimgov}-\ref{eq:nondim_mass_conservation}) is solved using a Fourier pseudo-spectral method \cite{Tan1988} and modified to account for Langmuir adsorption \cite{Rana2017a}. The number of spectral modes chosen for a computational domain of size $8192\times 1024$ is $2048 \times 256$. The spatial and time steps are taken as $\Delta x = \Delta y = 4$ and $\Delta t = 0.2$ respectively. Our code has been extensively tested against results from previous numerical simulations for a wide range of different flows \cite{Rana2014, Rana2017a, Rana2015}. 

\section{Interaction of Shock layer and Rarefaction wave:  $R=0$ case}
In order to study the interaction between the shock layer (SL) and the rarefaction wave (RF) formed due to Langmuir adsorption in the absence of VF ($R=0$), we plot the concentration fields at different times for $b=0$ and $b=2$ in Fig. \ref{fig:avg_b0b2_r0}(a). It is observed that, for $b=0$, both interfaces of the solute distribution show symmetrically diffusing profiles. However, for $b=2$ we have a highly diffused rear interface and a sharpened frontal interface, a characteristic concentration profile in presence of Langmuir adsorption \cite{Guiochon2006}. In order to investigate the difference between linear and Langmuir adsorption profiles of the solute, we compute the transverse averaged concentration profile defined as \cite{DeWit2005} :
\begin{equation}
\label{eq:trans_avg_cm} 
\bar{c}_m(x,t) = \frac{1}{\rm Pe}\int_0^{\rm Pe} c_m(x, y, t) {\rm d}y,
\end{equation}
and plotted in Fig. \ref{fig:avg_b0b2_r0}(b). The system is shown in a frame moving with the injection speed, so the solute is seen to move in the upstream direction. In Fig. \ref{fig:avg_b0b2_r0}, the concentration profile with $b=0$ presents the symmetric characteristics of a solute undergoing a linear adsorption, as already studied in detail previously \cite{Mishra2007}. For $b=2$, the Langmuir adsorption leads to the dependence of the migration rate on concentration, thus forming a rarefaction wave (RF) at the rear interface and a shock layer (SL) at the frontal interface. The shock-layer interface for $b=2$ is less dispersed in comparison to the frontal interface for $b=0$. On the other hand, the rarefaction wave widens the concentration distribution in comparison to the rear interface for $b=0$. The interaction of the rarefaction wave with the shock layer results in a decrease of the peak height. The concentration profile becomes nearly triangular and band broadening is observed to be enhanced in comparison to the linear adsorption case (see Fig. \ref{fig:avg_b0b2_r0}(b)). This kind of triangular concentration profile is observed in theoretical studies of chromatographic separation \cite{Lin1995} as well as in contaminant transport models \cite{BRUSSEAU1995} considering a non-linear adsorption of the solute. 

\begin{figure}[!htbp]
\begin{tabular}{c}
(a)\\
\includegraphics[width=0.45\textwidth]{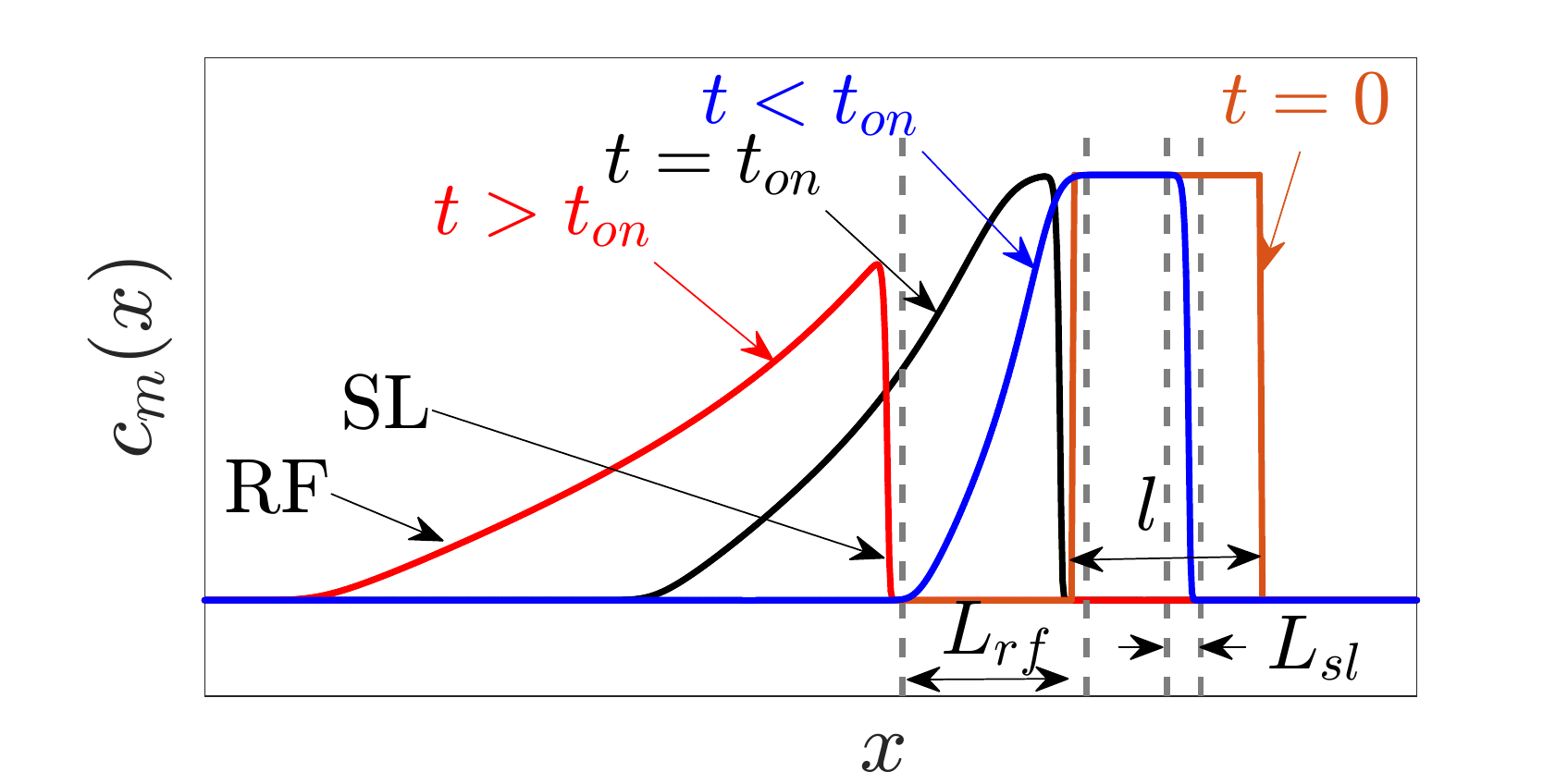} \\
(b)\\
\includegraphics[width=0.45\textwidth]{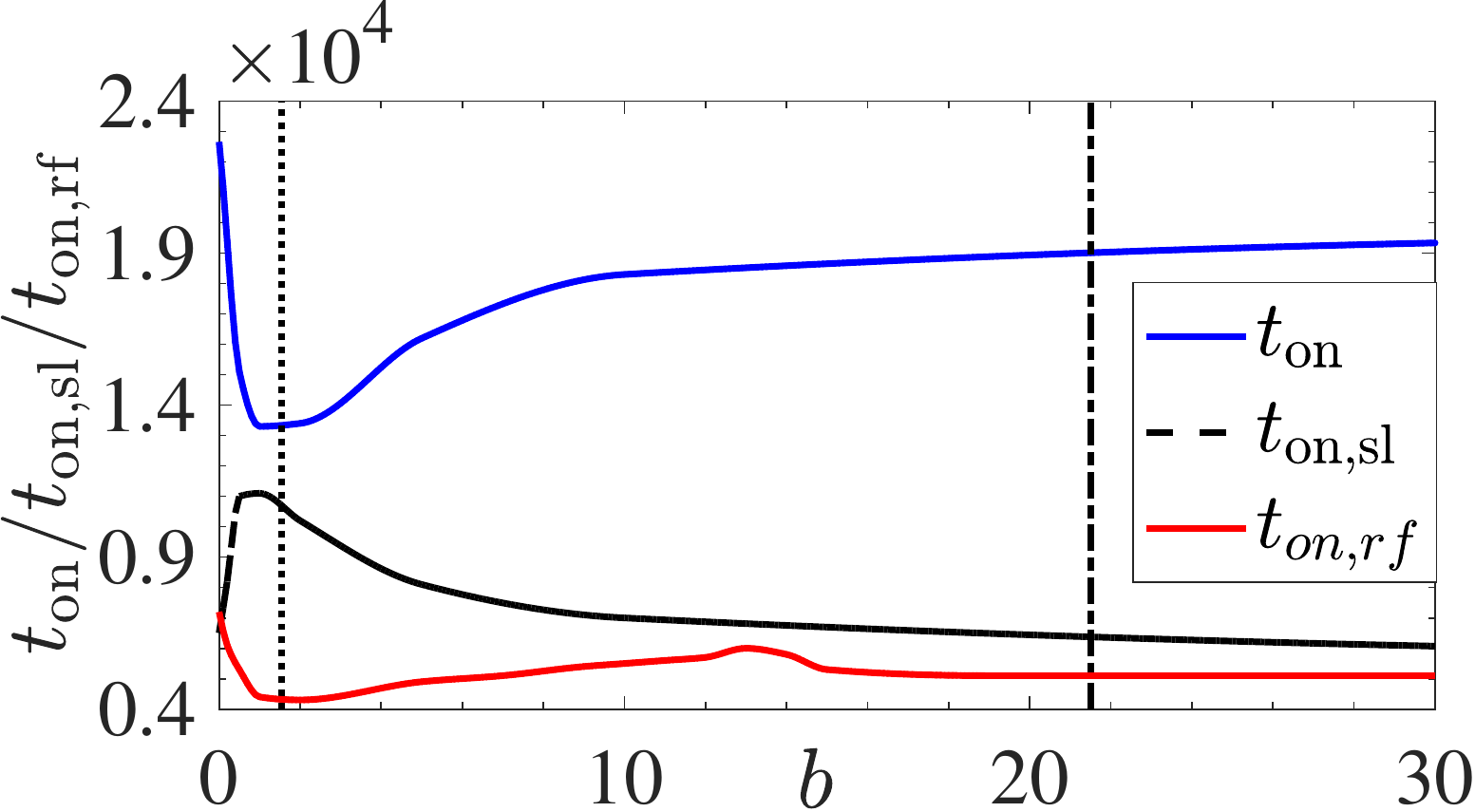}
\end{tabular}
\caption{(Color online) (a) 1D concentration profiles of a Langmuir adsorbed solute for $k=0.2,b=1$ at time $t=0$, $t<t_{\rm on}$ showing the definition of $L_{\rm rf}~ \text{and}~L_{\rm sl}$ (blue), $t=t_{\rm on}$ (time of onset of the interaction between SL and RF, black), $t>t_{\rm on}$ (long time triangle-like profile, red). (b) Interaction times $t_{\rm on}$ (SL with RF), $t_{\rm on,sl}$ (VF with SL) and $t_{\rm on,rf}$ (VF with RF) exhibit a non-monotonic dependence on $b$. Here $k=0.2$. The vertical lines correspond to $b^{\star}$ (dotted) and $b^{\dag}$ (dash-dotted).
} 
\label{fig:R0_cmx} 
\end{figure}

Since, for $R=0$ there is no instability in the transverse direction, $1$D concentration profiles capture all the dynamics of the solute. Thus, to analyze the interaction between SL and RF for $R=0$, we use a $1$D solute transport model for $c_m(x,t)$. In Fig. \ref{fig:R0_cmx}(a), we observe that, after a given time $t_{on}$, the SL and RF waves start to interact, the $c_m=1$ plateau does not exist any longer and a triangle-like profile is obtained. The time of this interaction $t_{\rm on}$ is defined as the time when: 
\begin{equation} \label{eq:ton}
 \max_{x} \{ c_m(x,t)\} < 1.
 \end{equation}
 The time of interaction between the shock layer and rarefaction waves depends on $l, k$ and $b$. For given values of $k$ and $b$, the onset time of interaction increases with $l$ while, for a fixed sample width $l$, $t_{on}$ depends non-trivially on $b$ and $k$. To show this, we first analyze separately the influence of these parameters on the width of SL and RF before analyzing the interaction between them (see Fig. \ref{fig:R0_cmx}(a)). 

\subsection{The Shock Layer thickness \label{subsec:SLT}} 

\begin{figure}
\centering
\includegraphics[width=0.45\textwidth]{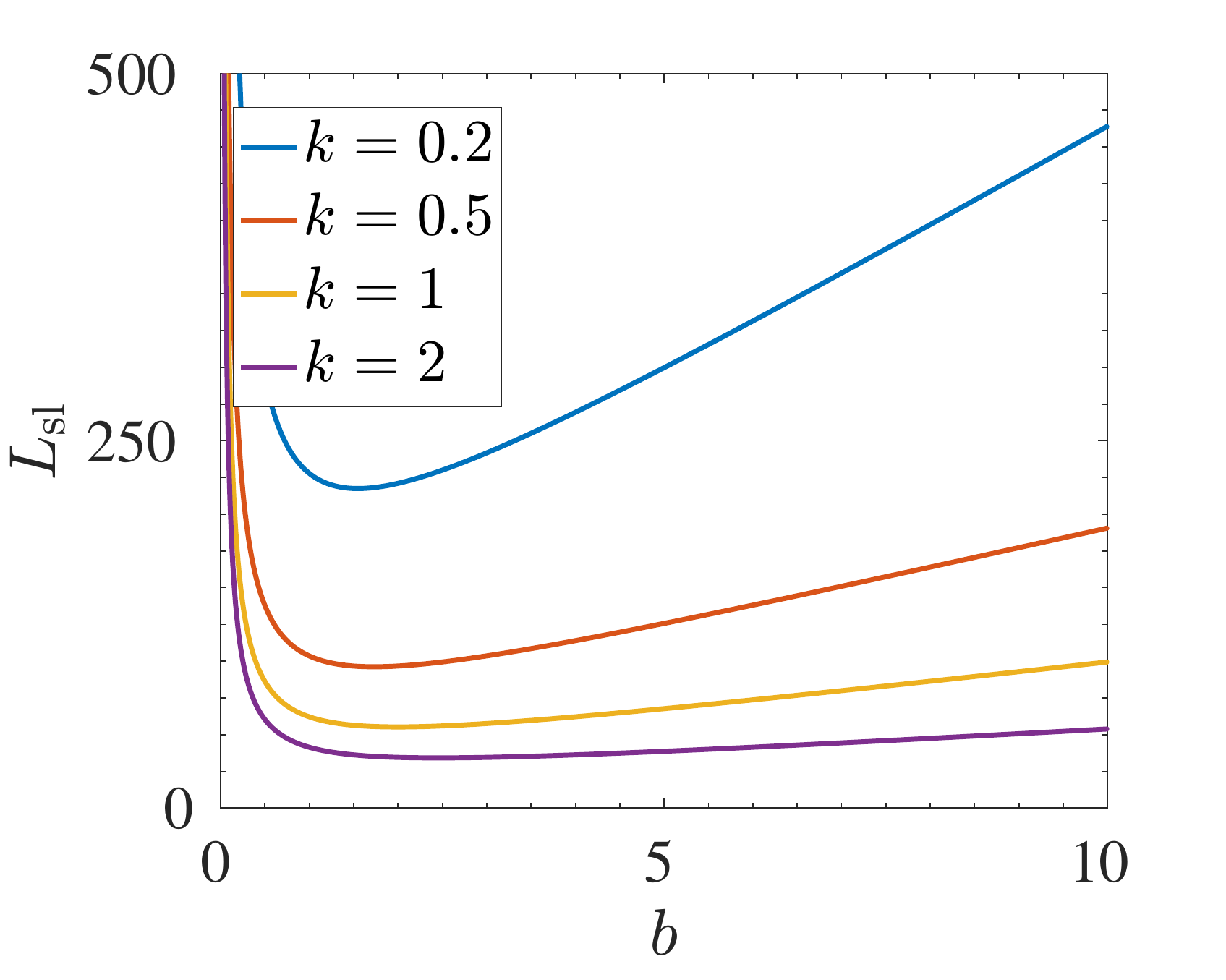}
\caption{Shock Layer thickness $L_{\rm sl}$ as a function of $b$ shows a global minimum at $b^{\star} = \sqrt{2(1+k)}$.} 
\label{fig:slt_vs_b}
\end{figure}

Following previous work (see Eq. (25) of \cite{Rana2017a}), we define the {\em shock layer thickness}, $L_{sl}$, as the width of the interval for which the concentration ${c}_m(x,t)$ at the frontal interface lies in the range $c_m^- < {c}_m(x, t) < c_m^+$: 
\begin{equation}
\label{eq:SLT}
L_{\rm sl} = \frac{(2+b)(1+k+b)}{kb}\ln\left(\frac{c_m^+}{c_m^-}\right), 
\end{equation}
which we rewrite as  $L_{\rm sl} = B~g(b;k)$, where $g(b;k) = (2+b)(1+k+b)/kb$ and  $B = \ln(c_m^+/c_m^-) \approx 6.91$ using $c_m^+ = 0.999$ and $c_m^- = 0.001$. From 
\begin{equation}
\label{eq:dgdb}
\frac{{d}g}{{d}b} = \frac{1}{k}\left[1 - \frac{2}{b^2}(1+k)\right] = 0 
\end{equation}
one obtains $b^{\star} = \sqrt{2(1+k)}$ (the negative root of the quadratic equation in $b$ is neglected owing to the fact that $b$ is positive). Further, 
\begin{equation}
\label{eq:d2gdb2}
\frac{{d^2}g}{{d}b^2}\Bigg\vert_{b^{\star}} = \frac{4}{b^{\star 3}}(1+k) = \sqrt{\frac{2}{1+k}} > 0 
\end{equation}
ensures that $g(b;k)$ has a global minimum at $b^{\star} = \sqrt{2(1+k)}$, and so does $L_{sl}$. In Fig. \ref{fig:slt_vs_b} we plot $L_{sl}$ as a function of $b$, from which it is clearly seen that $L_{sl}$ has a global minimum at $b^{\star} = \sqrt{2(1+k)}$. 
\subsection{The Rarefaction wave thickness \label{subsec:RFT}} 

Unlike the shock layer front, the rarefaction wave does not acquire a constant speed and shape \cite{Rana2017a}. In fact, the traveling speed of the solute varies locally with  its concentration $c_m$ in the mobile phase. Therefore, a closed form solution of the thickness (spreading length) of the rarefaction wave is not attainable. Here we present a crude approximation of the {\em rarefaction thickness}. In the absence of viscosity mismatch between the sample solvent and the displacing solution, the velocity of the solvent is a constant and in the dimensional form it takes the value of the injection velocity $U$. Therefore, in the moving coordinate frame, we have $\boldsymbol{\tilde{u}} = 0$ [c.f., Eqs. \eqref{eq:nondim_continuity}-\eqref{eq:nondim_mass_conservation}]. In this moving frame, the transport of $c_m$ is a resultant of two transport properties. One is the upstream advection at a concentration-dependent speed, $\mathcal{U}({c}_m)$ caused by the adsorption. The other one is dispersion with a concentration-dependent coefficient $\mathcal{D}({c}_m)$. Our argument is based on local approximations of the transport equation \eqref{eq:nondim_mass_conservation} at the rear interface in the neighborhood of ${c}_m(x,t) = 1 - \delta$ and ${c}_m(x,t) = \delta$ as 
\begin{subequations}
\begin{align}
& \partial_{t} {c}_m - \mathcal{U}_{1-\delta} \partial_x {c}_m = \mathcal{D}_{1-\delta}\partial_x^2 {c}_m, \label{eq:local_mass_conservation_1} \\
& \partial_{t} {c}_m - \mathcal{U}_{\delta} \partial_x {c}_m = \mathcal{D}_{\delta}\partial_x^2 {c}_m, \label{eq:local_mass_conservation_2} 
\end{align}
\end{subequations}
respectively. These are advection-diffusion equations with constant transport coefficients. Denoting the positions of ${c}_m = 1-\delta$ and ${c}_m = \delta$ at time $t$, as $x_{1-\delta}(t)$ and $x_{\delta}(t)$ respectively, we compute 
\begin{subequations}
\begin{align}
& x_{1-\delta}(t) = x_{1-\delta}(0) - \mathcal{U}_{1-\delta} t + \sqrt{2 \mathcal{D}_{1-\delta}t}, \label{eq:xR} \\ 
& x_{\delta}(t) = x_{\delta}(0) - \mathcal{U}_{\delta} t - \sqrt{2 \mathcal{D}_{\delta}t}. \label{eq:xL}
\end{align}
\end{subequations}
We define the spreading length of the rarefaction wave $L_{rf}$ as the width of the interval for which the concentration ${c}_m(x,t)$ at the rear interface lies in the range $\delta < {c}_m(x, t) < 1- \delta$: 
\begin{equation}
\label{eq:rarefaction_thickness}
L_{\rm rf}(t; b, k) = x_{1-\delta}(t) - x_{\delta}(t), 
\end{equation}
which, in combination with Eqs. \eqref{eq:xR} and \eqref{eq:xL} becomes 

\begin{figure}[!hbtp]
\includegraphics[width=0.45\textwidth]{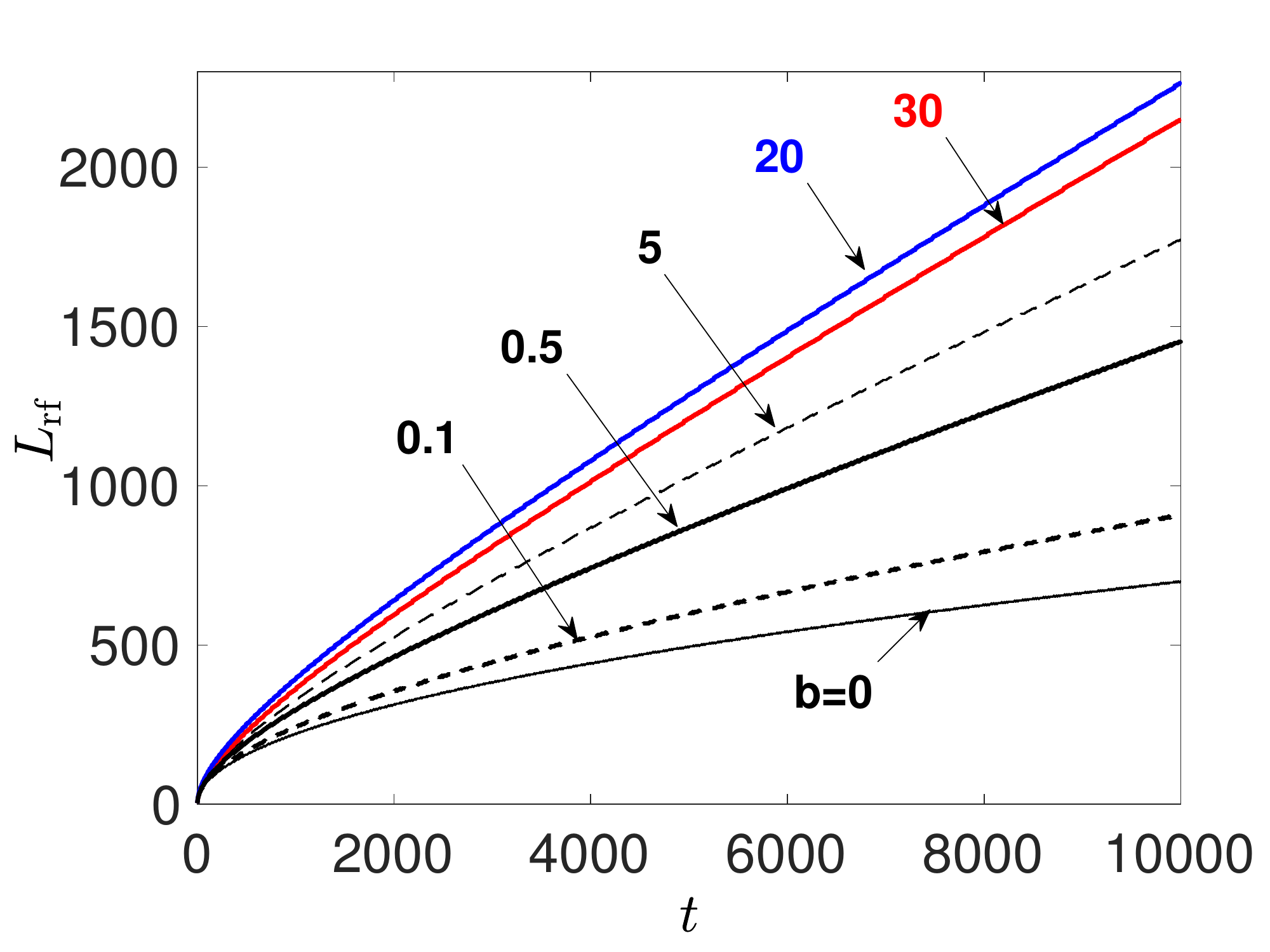}
\caption{The temporal evolution of the rarefaction thickness $L_{\rm rf}$ for different $b$ with $k=0.2$ shows a decrease in $L_{\rm rf}$ beyond $b=20$.} 
\label{fig:Lrf_r0} 
\end{figure}

\begin{widetext}
\begin{eqnarray}
L_{\rm rf}(t;b,k) - L_{\rm rf}(0;b,k) = \left[\frac{(1 + \beta b)}{\sqrt{k + (1 + \beta b)^2}} + \frac{(1 + \alpha b)}{\sqrt{k + (1 + \alpha b)^2}}\right]\sqrt{2 t}\nonumber \\+ \left[\frac{1}{k + (1 + \alpha b)^2} - \frac{1}{k + (1 + \beta b)^2}\right]k t, 
\end{eqnarray} 
where $\alpha = \delta$ and $\beta = 1-\delta$. 
For a fixed $k = k_0$ and $\forall t > 0$, we find $b^{\dag}(k_0)$ that maximizes $L_{\rm rf}$ computing 
\begin{eqnarray} 
\label{eq:deriv_Lrf}
\frac{{\rm d}L_{\rm rf}}{{\rm d}b} = \left[ \frac{ \beta }{\left[ k + (1 + \beta b)^2 \right]^{3/2} } + \frac{ \alpha }{\left[ k + (1 + \alpha b)^2 \right]^{3/2} } \right] \sqrt{2 t k^2} \nonumber \\+ \left[ \frac{ \beta (1 + \beta b)}{\left[ k + (1 + \beta b)^2 \right]^{2} } - \frac{ \alpha (1 + \alpha b)}{\left[ k + (1 + \alpha b)^2 \right]^{2} } \right] 2 k t, 
\end{eqnarray}
which, equating to zero yields $b^{\dag}(k_0)$. Further, 
\begin{eqnarray}
 \frac{ {\rm d^2} L_{\rm rf} }{ {\rm d} b^2} \Bigg\vert_{b^{\dag}} = \left[ \frac{ -3 \beta^2 (1 + \beta b^{\dag}) }{ \left[ k + (1 + \beta b^{\dag})^2 \right]^{5/2} } + \frac{ -3 \alpha^2 (1 + \alpha b^{\dag}) }{ \left[ k + (1 + \alpha b^{\dag})^2 \right]^{5/2} } \right] \sqrt{2 t k^2} \nonumber \\ 
\label{eq:deriv2_Lrf} 
 \qquad \qquad + \left[ \frac{ \beta^2 \left[ k - 3 (1 + \beta b^{\dag})^2 \right] }{\left[ k + (1 + \beta b^{\dag})^2 \right]^{3} } - \frac{ \alpha^2 \left[ k - 3 (1 + \alpha b^{\dag})^2 \right] }{ \left[ k + (1 + \alpha b^{\dag})^2 \right]^{3} } \right] 2 k t < 0. 
\end{eqnarray}
\end{widetext}

For a given value of $k$, we obtain $b^{\dag}(k)$ such that $L_{\rm rf}[t; b^{\dag},k] > L_{\rm rf}[t; b, k], ~ \forall b \neq b^{\dag}$ and $\forall t > 0$. Thus the rarefaction thickness $L_{\rm rf}$ increases with $b$ until the widest RF wave is achieved for a particular $b^\dag$ depending on $k$. We compute $b^{\dag}(k = 0.1) \approx 19.3116$, $b^{\dag}(k = 0.2) \approx 21.4672$, and $b^{\dag}(k = 1) \approx 62.2245$. In Fig. \ref{fig:Lrf_r0} the thickness of RF wave front $L_{\rm rf}$ is plotted for $k=0.2$ and different values of $b$. Clearly, $L_{\rm rf}$ increases for $0<b<20$ and decreases for $b=30$, justifying the above calculation where $b^ \dag \simeq 21$ for $k=0.2$.  

\begin{figure*}
\begin{tabular}{cc}
 \hspace{-2.7in}(a) &  \hspace{-3in}(b)\\
 \hspace{-2.7in} \includegraphics[scale=0.4]{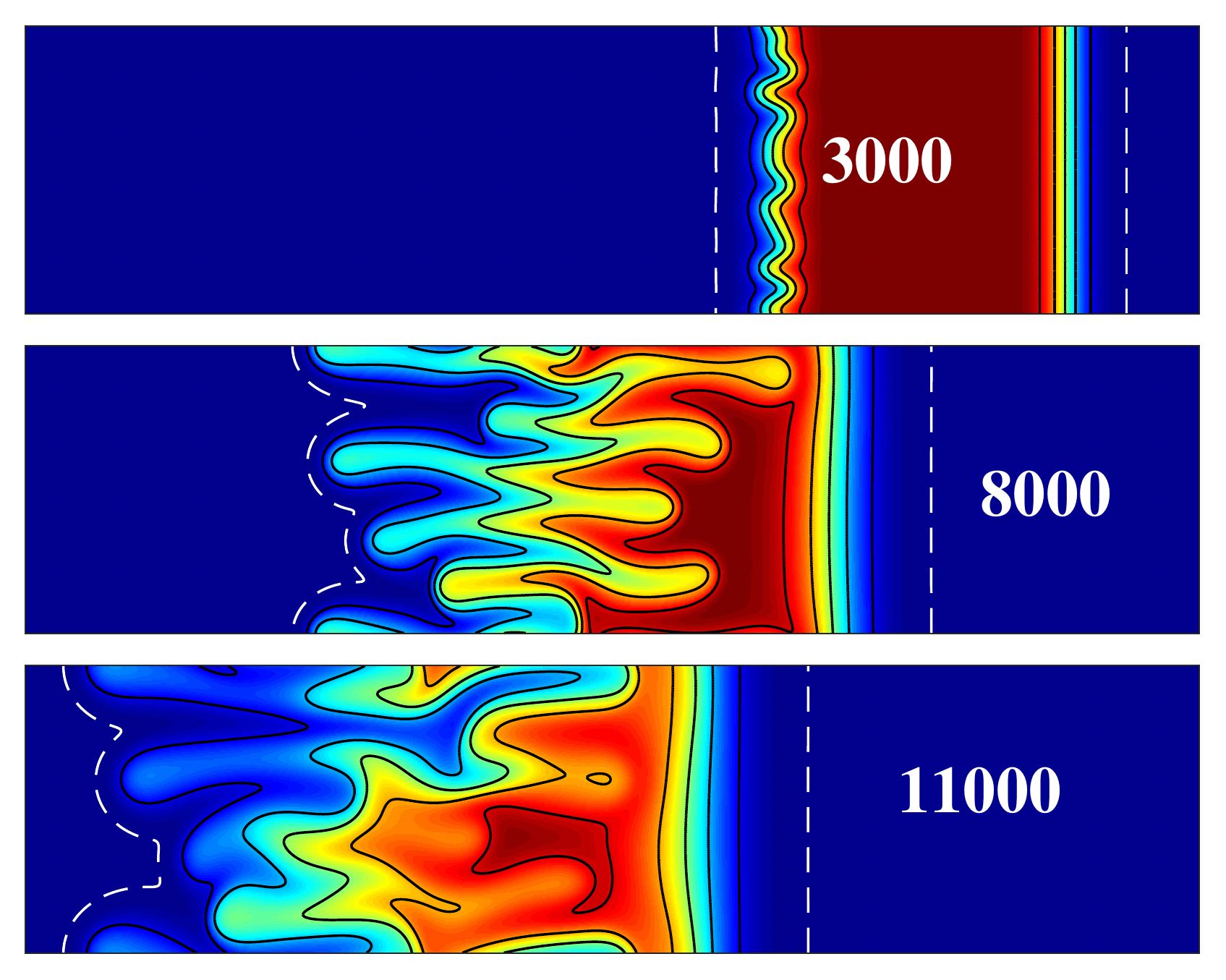} & \hspace{-2.7in}
\includegraphics[scale=0.4]{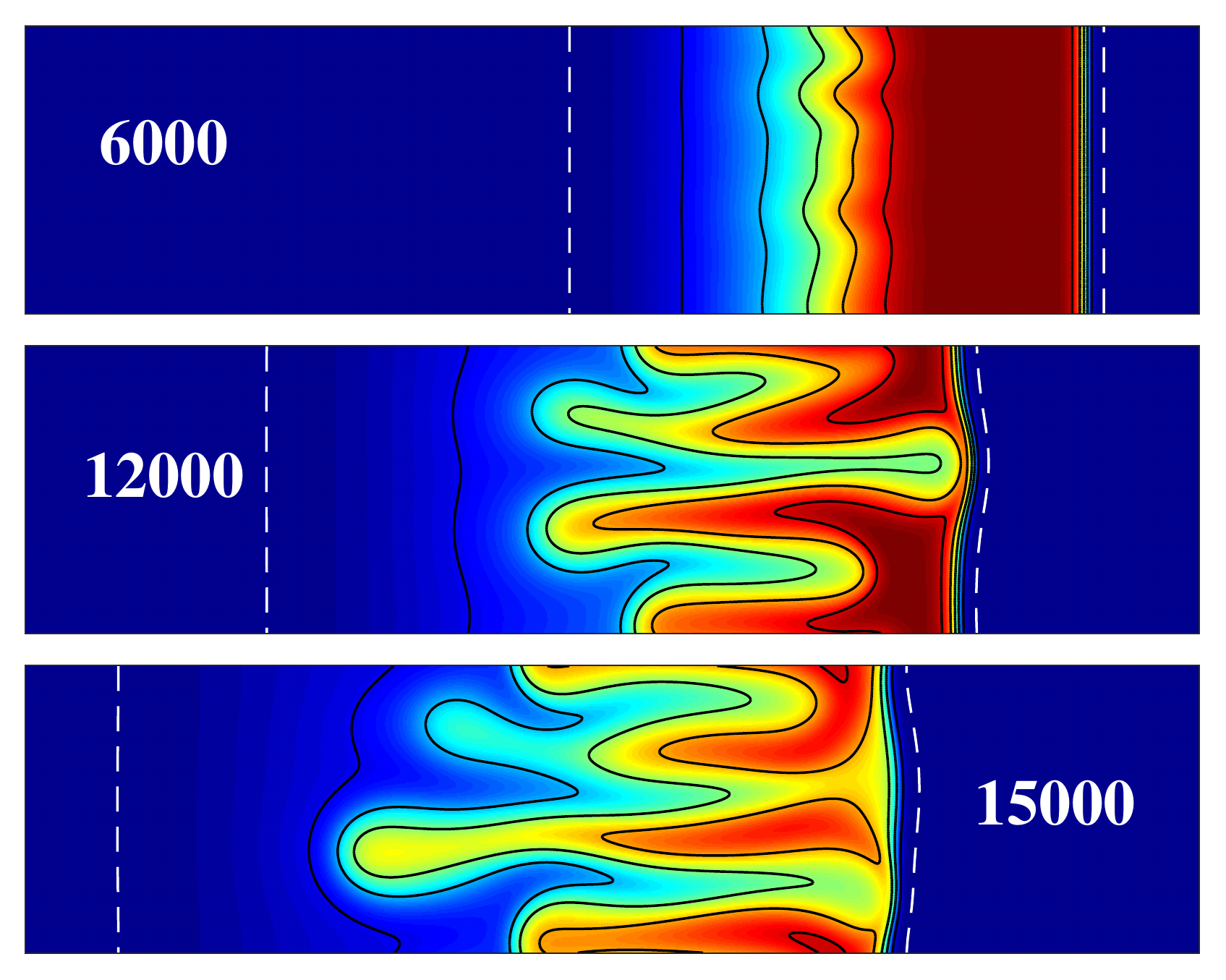}\\
 \hspace{-2.7in} (c) &  \hspace{-2.7in} (d)\\
\includegraphics[scale=0.4]{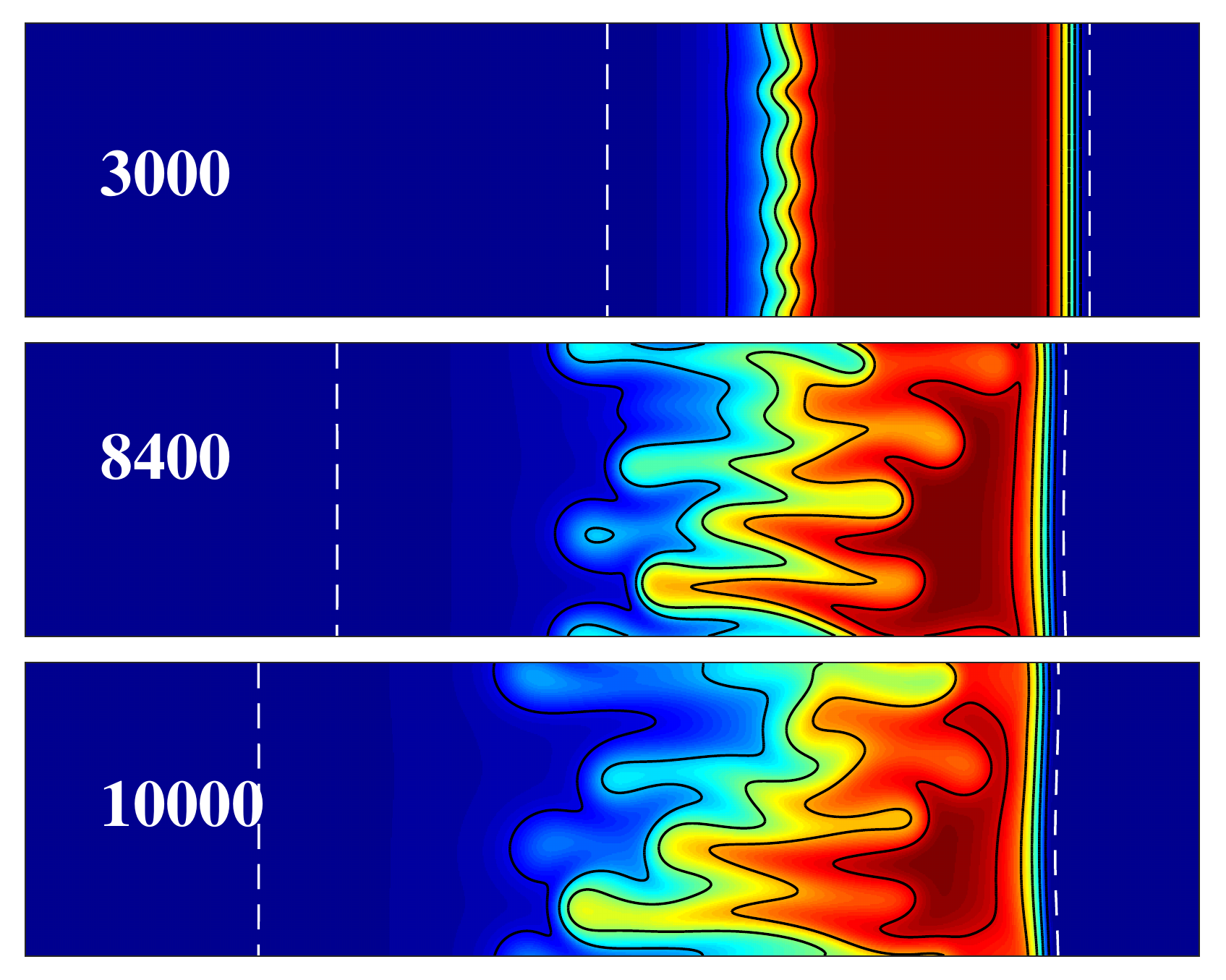}
\includegraphics[scale=0.4]{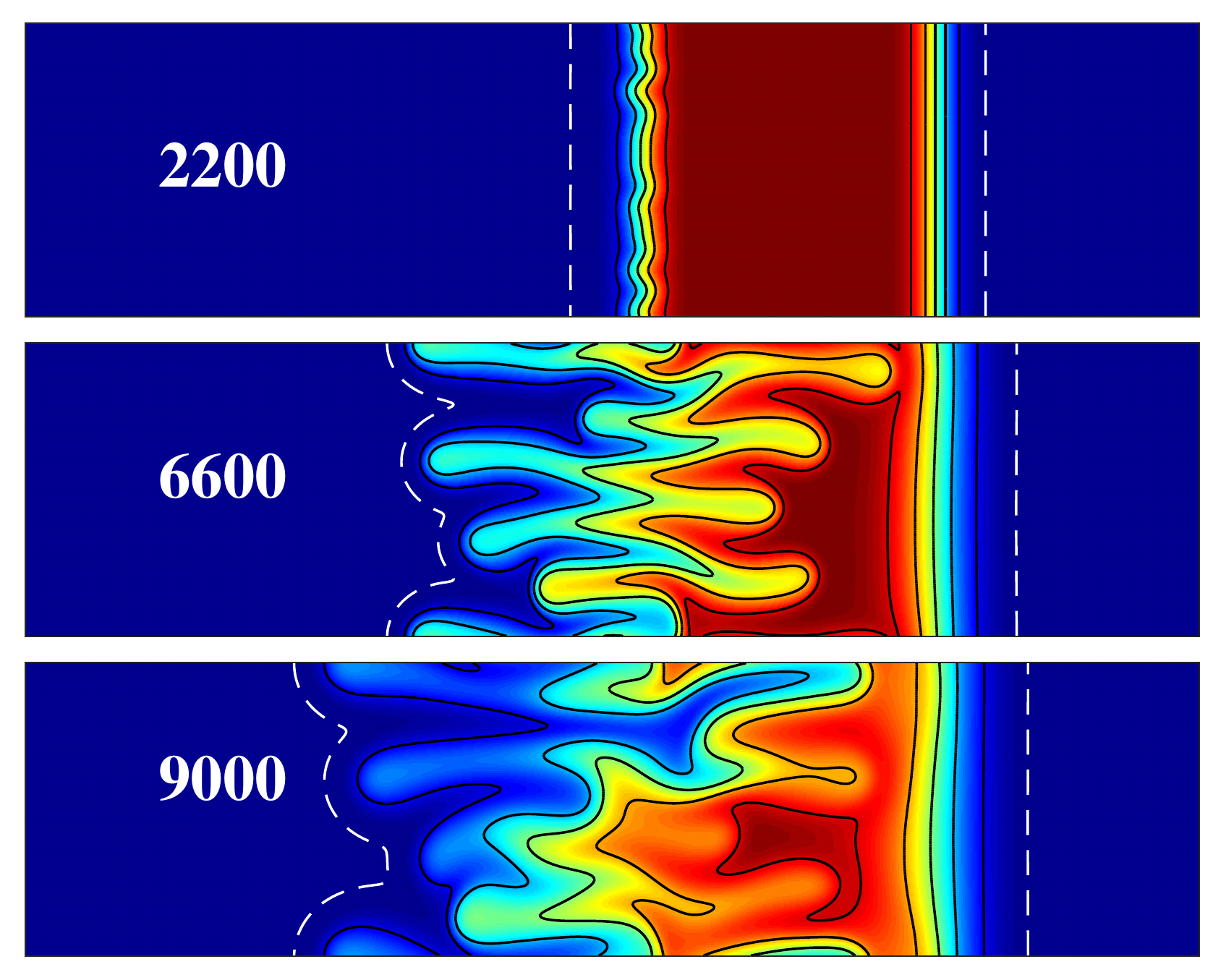}\\
\end{tabular}
\caption{(Color online) Concentration maps for $R=1, k = 0.2$: (a) $b$ = 0, (b) 1.55 ($=b^{\star}$), (c) 10 and (d) $10^3$ (the saturated case) at the time written in the panel. The solid contours are $c_m = 0.1$ to 0.9 with step 0.2. The dashed contour corresponds to $c_m = 0.001$. The concentration maps are shown at the onset of VF (top row), at $t \approx t_{\rm on, sl}$ (middle row), and at $t > t_{\rm on, sl}$.
} 
\label{fig:R1} 
\end{figure*}

\subsection{Evolution of interaction time $t_{on}$ with $b$ }
The time of interaction between SL and RF, $t_{on}$, is evaluated as the time at which the relation defined in Eq.\ref{eq:ton} is satisfied and plotted in Fig. \ref{fig:R0_cmx}(b) as a function of $b$. Clearly, $t_{on}$ is observed to vary non-monotonically with $b$. For small values of $b$ ($b\leq 1$), $t_{on}$ decreases rapidly and then starts increasing for $b>1$. The occurrence of this non-monotonicity in $t_{on}$ can be explained as follows: For a very large value of $b$ ($b \to \infty$), we have the saturated case where the solute molecules in the mobile phase migrate nearly without interacting with the stationary phase, so that the SL ceases to exist and the $c_m(x)$ profile approaches an error function. Thus $L_{\rm sl}$ has a non-monotonic dependence on $b$ (see Fig. \ref{fig:slt_vs_b}). For the RF wave, the spreading length is widest for $b=b^\dag$ thus $L_{\rm rf}$ increases for $b< b^{\dag}$. As mentioned above, we recover the error function profile in the limit $b \rightarrow \infty$, hence for large $b$, $L_{\rm rf} \rightarrow l_{d}$, the spreading length of a diffusive front ($\propto \sqrt{t}$). Thus, $L_{\rm rf}$ first increases and then decreases as $b$ increases. Also, as studied previously \cite{Rana2015}, for a given $k$ and at a time $t$, we obtain $x({c}_m = 0.999, t, k; b_i) < x(	{c}_m = 0.999, t, k; b_j)$, $b_i < b_j$ at the RF front. Therefore, an increase in $b$ shortens the distance between the apex of the RF and SL waves. This, in combination with a non-monotonic $L_{\rm sl}$ and $L_{\rm rf}$, causes a non-monotonic dependence of $t_{\rm on}$ on $b$.  Now that we have characterized the interaction between SL and RF waves, let us analyze how they evolve in presence of VF.

\section{Interaction of nonlinear waves with viscous fingering}
\subsection{Interaction between SL and VF for $R>0$}
For $R>0$, the rear interface of the finite slice of sample becomes unstable and shows viscous fingering. With time, the fingers develop and interact with the stable SL frontal interface of the Langmuir adsorbed solute. Following the interaction between the fingers and the SL interface, the plateau $\bar{c}_m=1$ disappears and the maximum of $\bar{c}_m$ starts decreasing. Thus to quantify the onset of interaction between viscous fingering and SL, we compute $t_{\rm on, sl}$ as the time at which $\displaystyle \max_{x} \{ \overline{c}_m(x,t)\} < 1$ (see Fig. \ref{fig:R0_cmx}(b)). It is known that a linear adsorption ($b = 0$) delays the onset of fingering compared to the saturated case ($b \to \infty$) \cite{Mishra2007}. For Langmuir adsorption and intermediate values of $b$ ($0 < b < \infty$), the onset time of VF varies non-monotonically with $b$ as shown in Figs. \ref{fig:R1}(a-d). Accordingly a non-monotonic dependence of $t_{\rm on, sl}$ on $b$ is obtained [see Figs. \ref{fig:R0_cmx}(b) and \ref{fig:R1}(a-d)]. In addition, we observe that the SL front distorts because of the interaction with the fingers. As, in a SL, the concentration profiles should all propagate with the shock velocity such that the concentration contours are straight lines (see Figure \ref{fig:R1}(b) at time $t = 6000$), we do not have stricto sensu a SL anymore once it is affected by fingering. After the interaction with the fingers, the steep stable viscosity contrast at the frontal part blunts the forward moving fingers that try to intrude in the SL [Figs. \ref{fig:R1}(b-c)].

\begin{figure*}
\begin{tabular}{ccc}
(a)  & (b) \\
\includegraphics[scale=0.3]{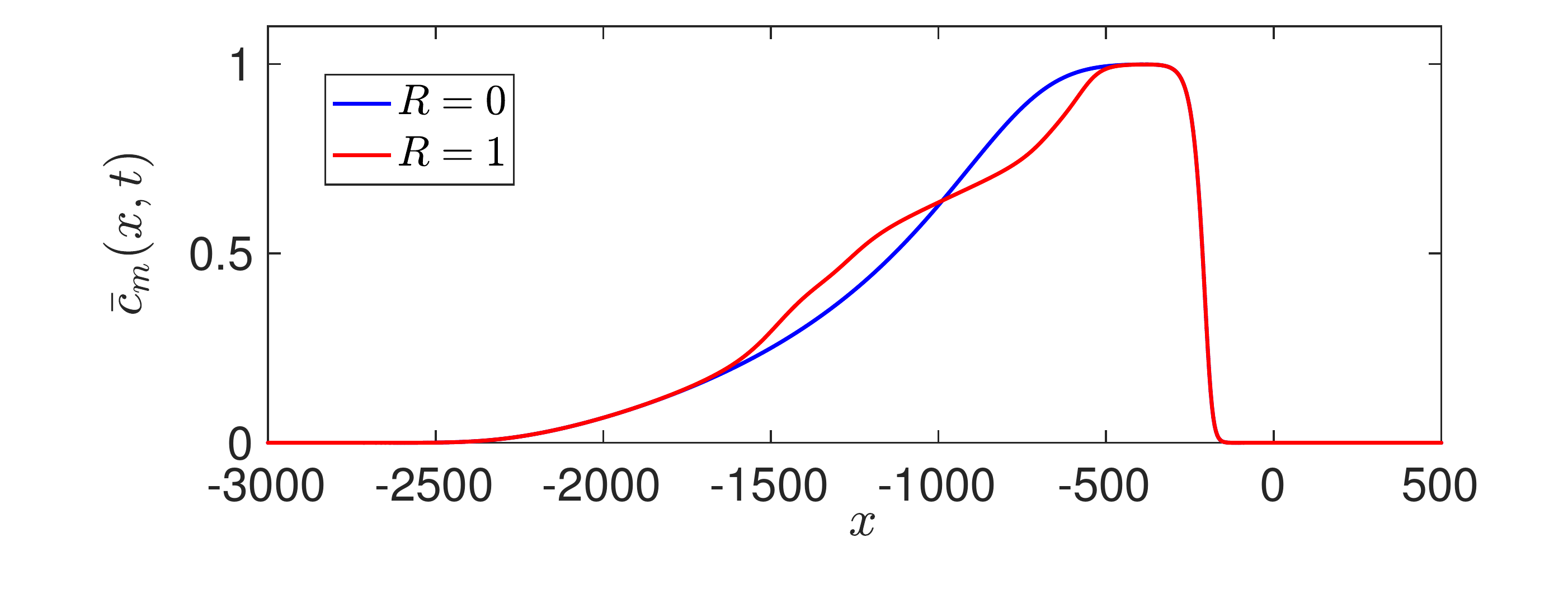} &
\includegraphics[scale=0.3]{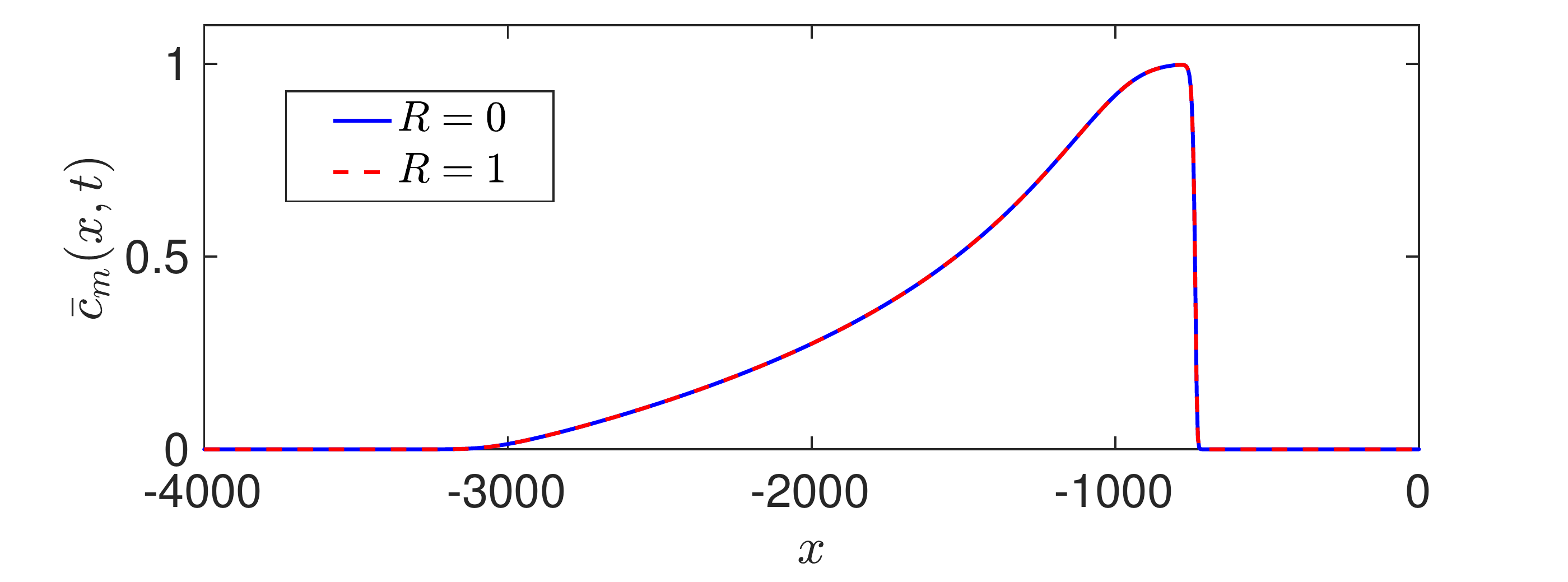}\\
(c) & (d)\\
\includegraphics[scale=0.3]{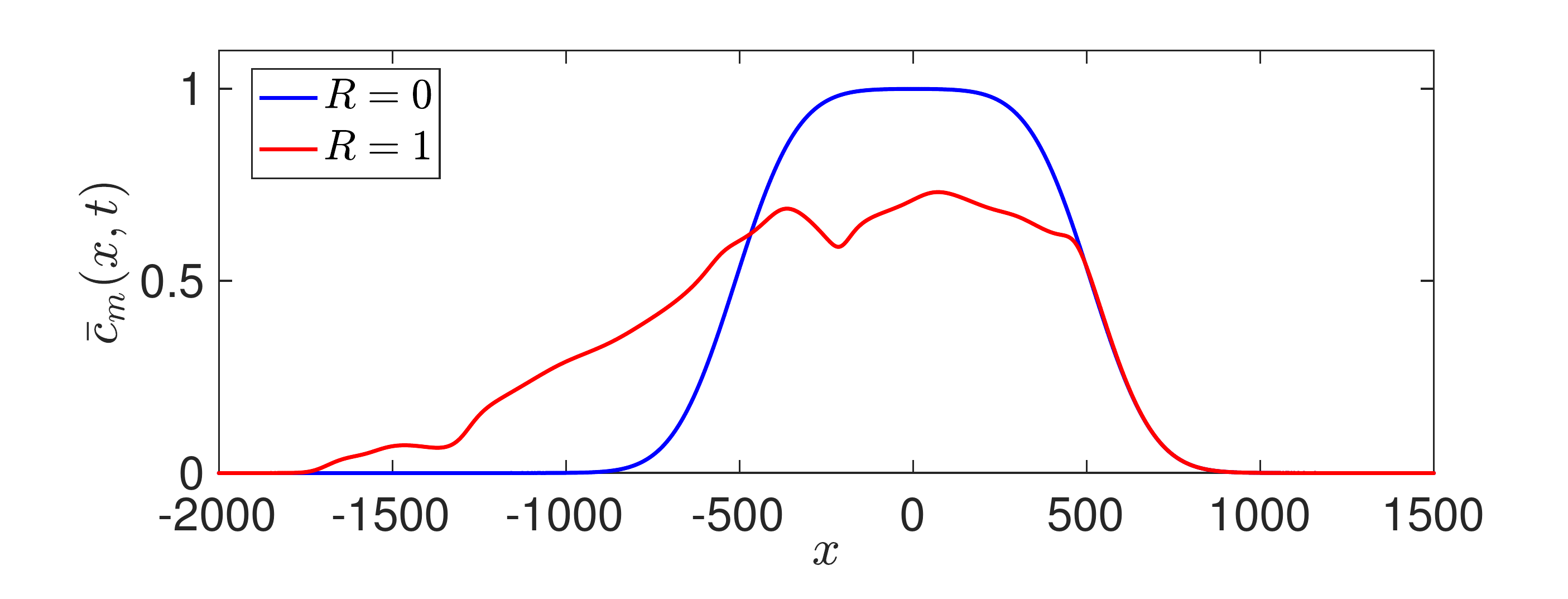} & 
\includegraphics[scale=0.3]{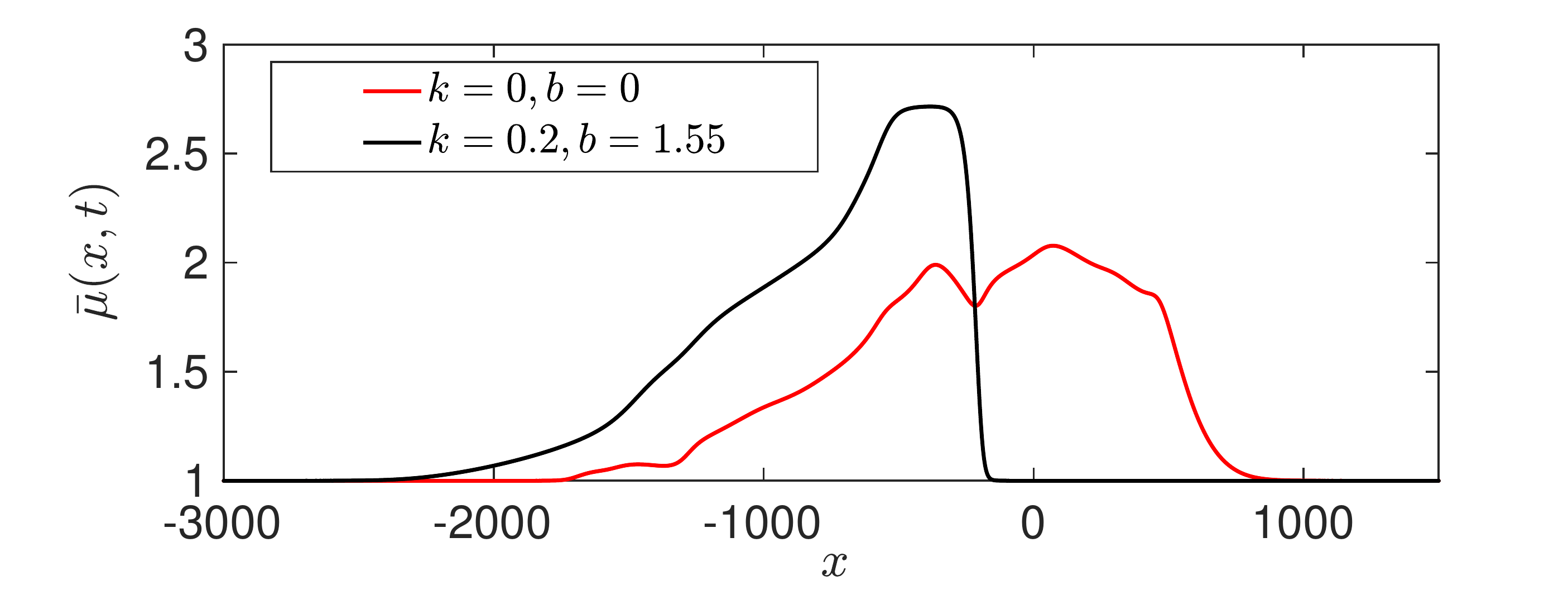}
\end{tabular}
\caption{Comparison of transverse averaged concentration profiles $\bar{c}_m(x)$ of the solute at time $t=10000$ for (a) $k=0.2,b=1.55$, (b) $k=1, b=2$, (c) $k=0,b=0$ i.e. no-adsorption case and $R=1$ or $R=0$. (d) Transverse averaged viscosity profiles $\bar\mu(x,t)$ for no-adsorption ($k=0,b=0$) and Langmuir adsorption ($k=0.2,b=1.55$) with $R=1$ at time $t=10000$. } 
\label{fig:avgconc} 
\end{figure*}

Figs. \ref{fig:avgconc}(a) and \ref{fig:avgconc}(b) show the transverse averaged concentration, $\bar{c}_m(x)$ at $t = 10000$ for different values of $k$ and $b$. For increased adsorption ($k = 1, b = b^{\star}(= 2)$), we see that viscous fingering can be significantly delayed or completely suppressed by Langmuir adsorption [see Fig. \ref{fig:avgconc}(b)]. Therefore, for certain combination of dimensionless parameters $R, \;b, \; k$, the effect of viscosity contrast can even be totally slaved to the Langmuir adsorption. Interestingly, the VF-induced distorted solute profile remains confined within the spreading zone of the viscously stable case (as in Fig. \ref{fig:avgconc}(a)). This is different from the situation of no-adsorption or linear adsorption \cite{DeWit2005, Mishra2007}, where the unstable displacement front spreads out of the region covered during the stable displacement (as in Fig. \ref{fig:avgconc}(c)).

The physical reason for the confinement of the fingering with Langmuir adsorption ($R=1$) in the region covered in the absence of VF ($R=0$) is the following. From the viscosity profile shown in Fig. \ref{fig:avgconc}(d), it is clear that for Langmuir adsorption, the viscosity gradients at the unstable rear interface differ significantly from those corresponding to no-adsorption. In the former case, the viscosity gradients are such that the fingers form near the high concentration zone (i.e. sufficiently away from the $c_m=0.001$ contour, see Figs. \ref{fig:R1}(b) and \ref{fig:R1}(c)). After the interaction with the stable frontal interface, the fingers reorient and propagate in the upstream direction. We remind from Eq.\eqref{eq:nondim_mass_conservation} that, in addition to being advected by the flow $\bu$, the adsorbed solute has an additional upstream advection speed $\mathcal{U}(c_m)\bex$ that increases with decreasing $c_m$ (see Eq.\eqref{eq:U}). Hence, the upstream moving tip of the fingers have a lower speed than the lowest concentration zone of the tail of the RF front. As an example, see in Fig. \ref{fig:R1}(b) that the $c_m=0.001$ contour which has a lower concentration than the tip of the fingered zone, travels faster upstream. As a consequence, the fingers are never able to catch up the upstream movement of the RF tail and fingering remains contained in the region covered by the solute in absence of fingering. Thus the decisive role played by the Langmuir adsorption is that, even when the dynamics are under the influence of VF, the solute transport cannot extend beyond the zone delimited by the purely diffusive regime.

These findings have important applications in chromatography separation. In typical chromatographic conditions, the viscosity of the sample is larger than that of carrier fluid. For a given solvent and solute, i.e. for a fixed value of the control parameter $R$, one can fix the nature of the porous matrix to a desired saturation capacity and saturation rate such that either VF does not occur at all in the column or VF induced distorted peaks of the solute are confined within the spread of the triangle-like waves that are formed in the absence of VF. In addition, the solvent injection rate can be controlled in such a manner that the solute travels the entire column length before an interaction of the fingers with the stable shock layer front reorients the former in the upstream direction to enhance the spreading over a larger column length. 
\begin{figure*}
\centering
\begin{tabular}{cc}
(a) &(b) \\
\includegraphics[scale=0.4]{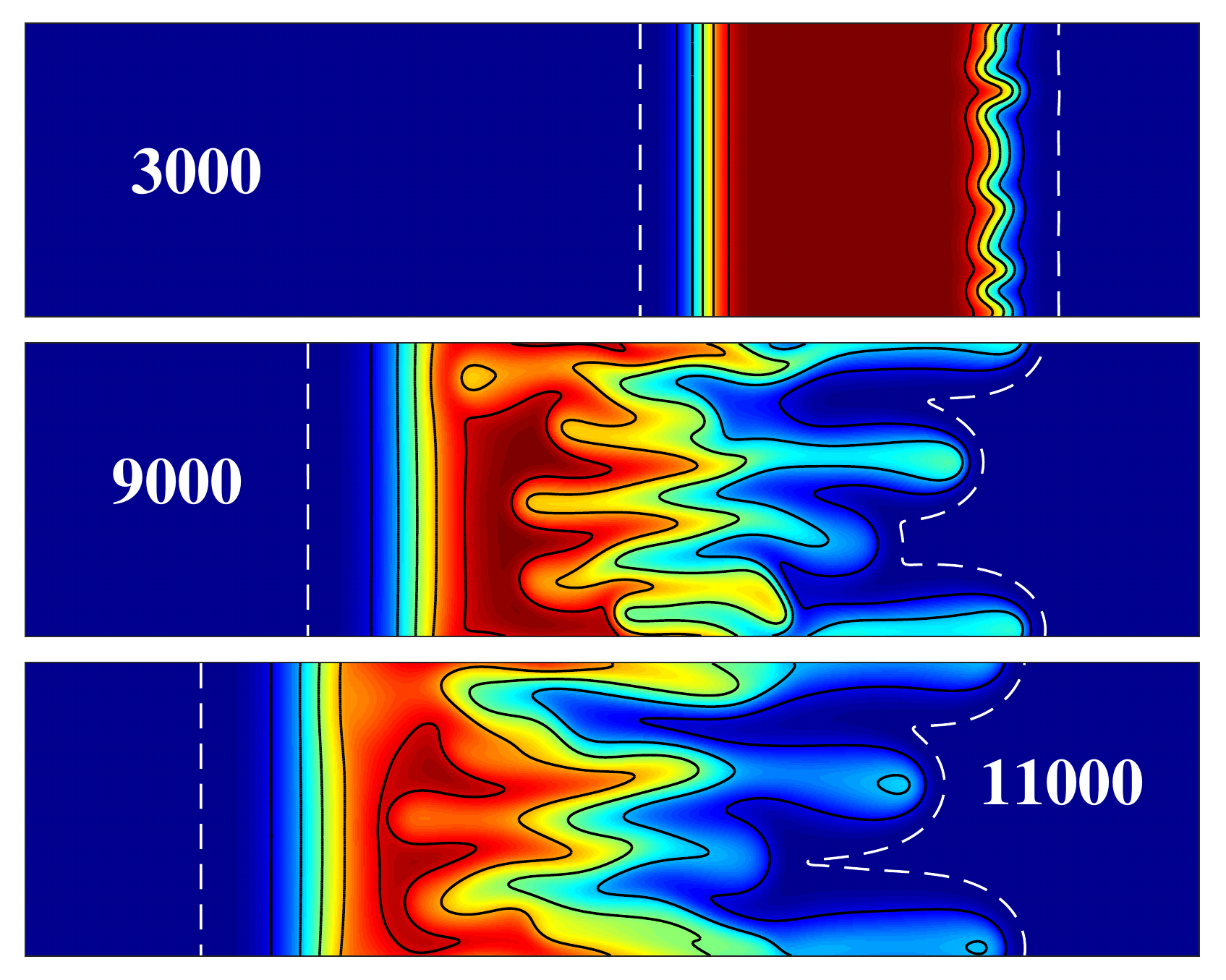} &
\includegraphics[scale=0.4]{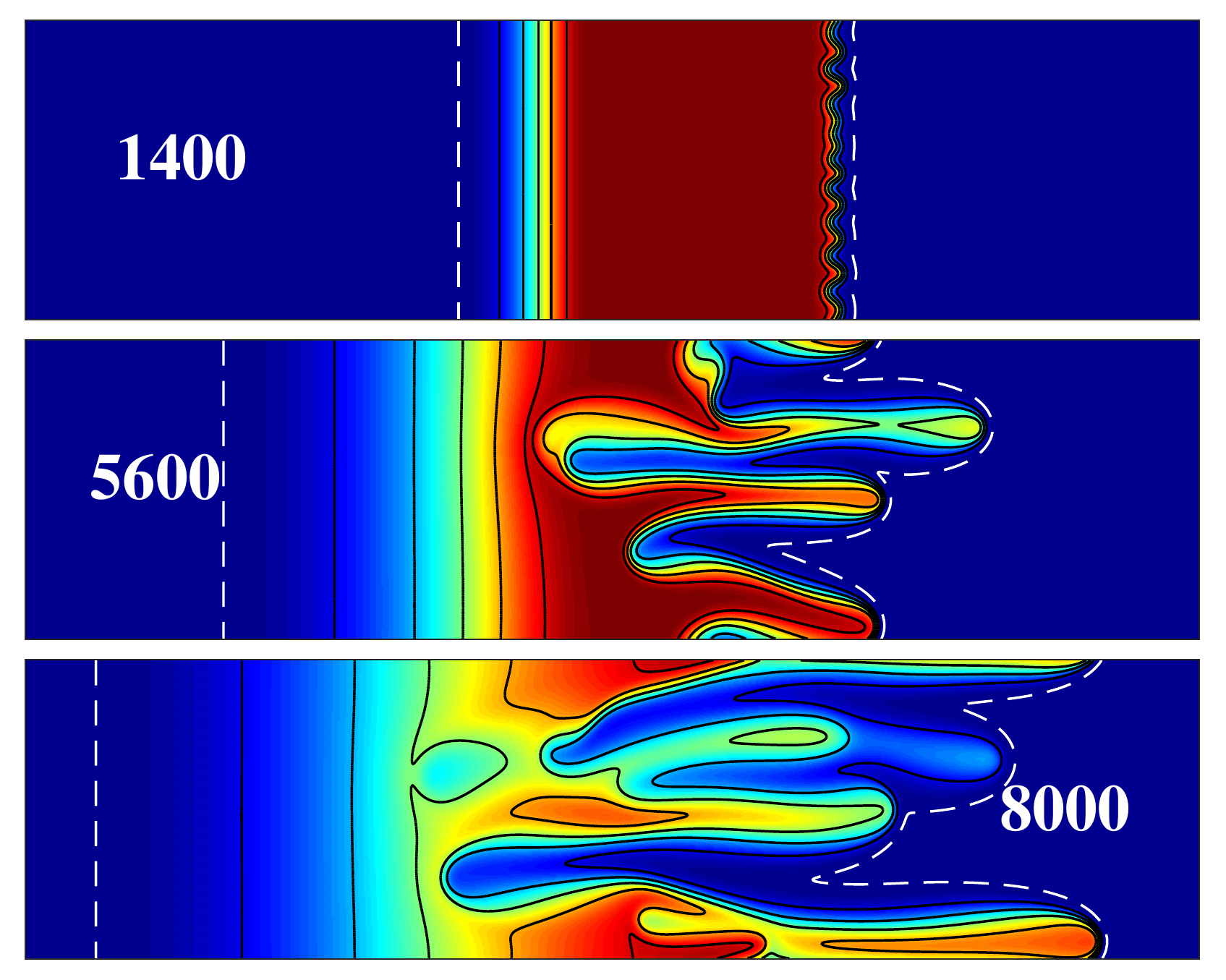}\\
(c) &  (d)\\
\includegraphics[scale=0.4]{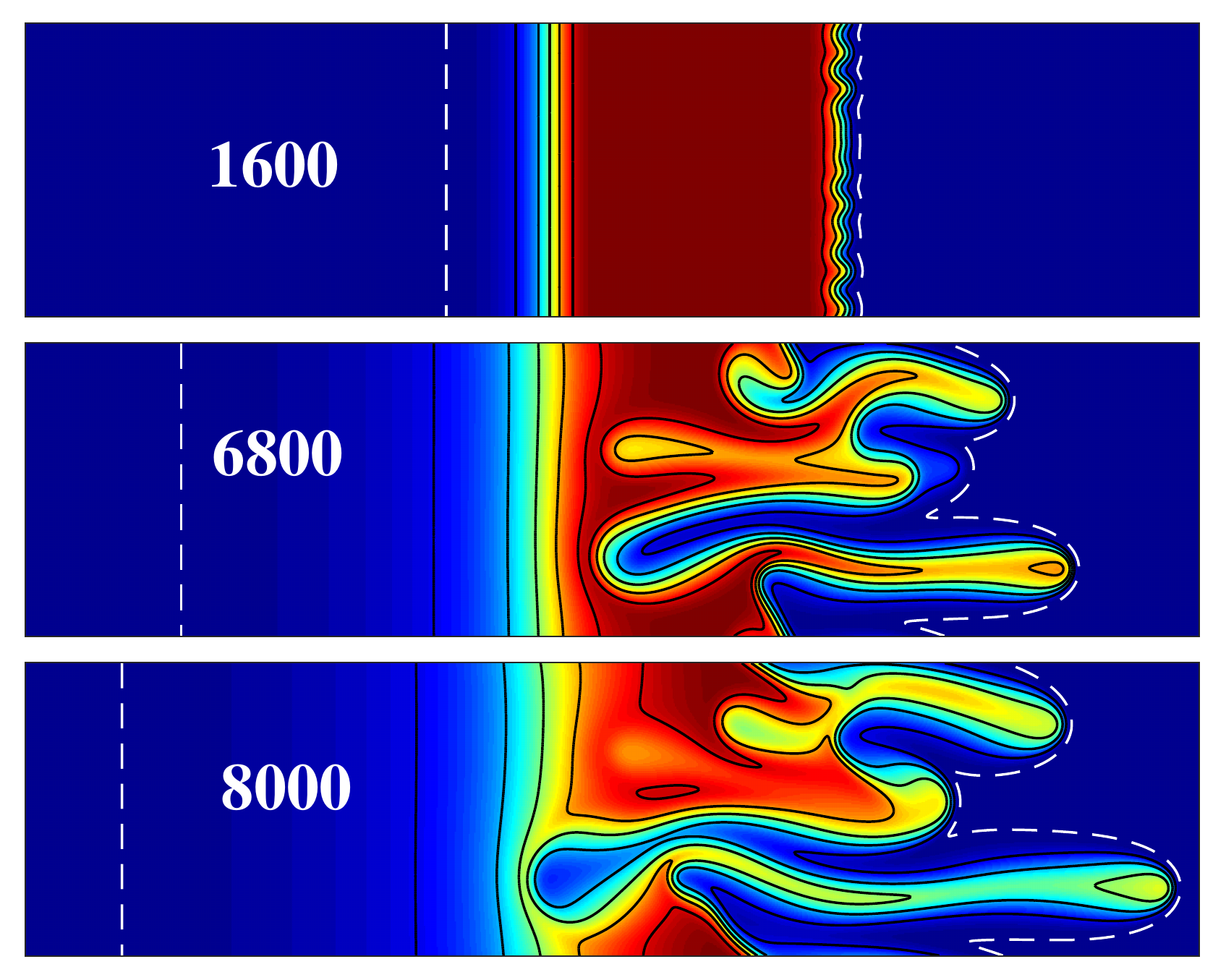} &
\includegraphics[scale=0.4]{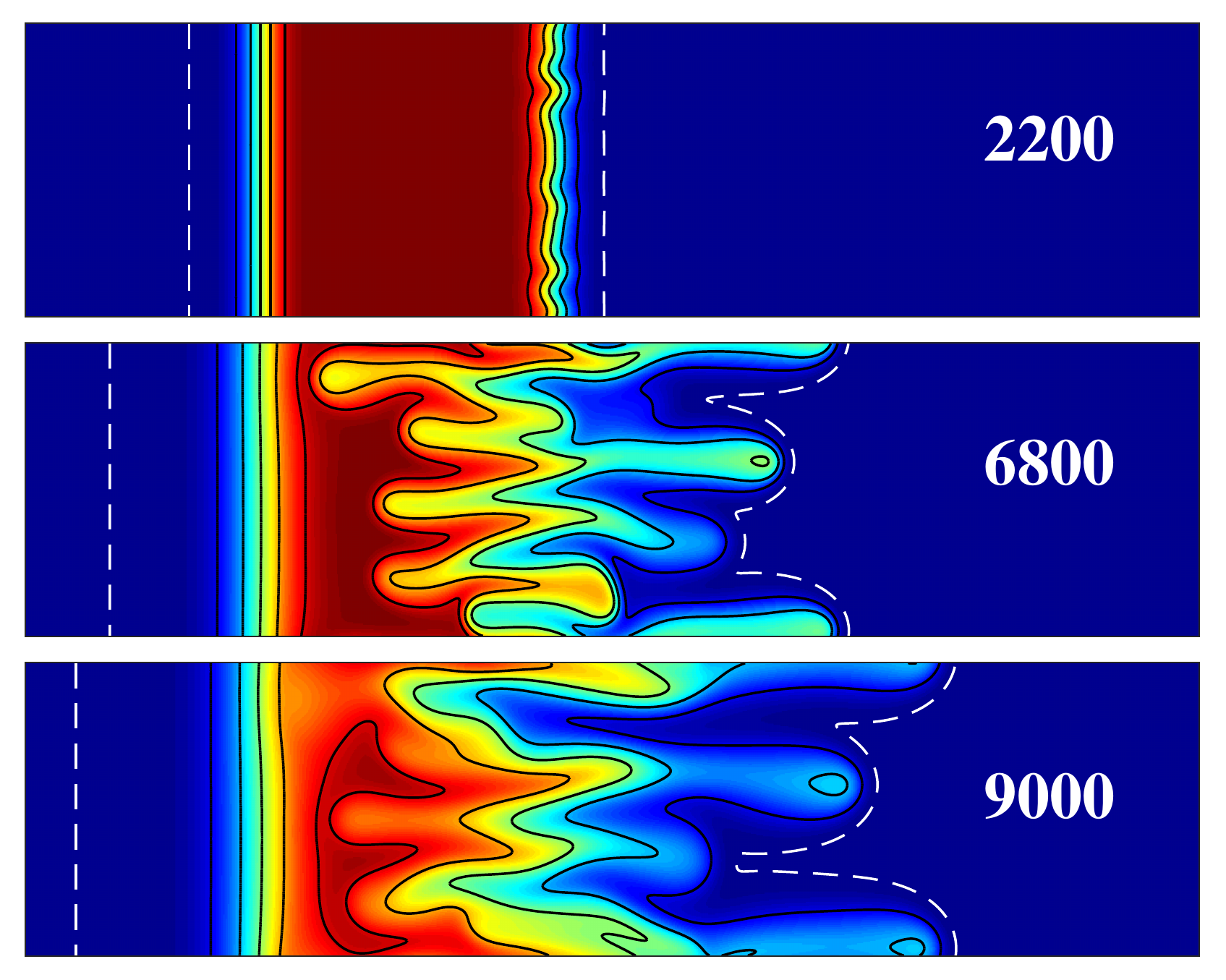}
\end{tabular}
\caption{(Color online) Concentration maps for $R = -1,k = 0.2$: (a) $b$ = 0, (b) 1.55 ($=b^{\star}$), (c) 10 and (d) $10^3$ (the saturated case) at the time written in the panel. The solid contours are $c_m = 0.1$ to 0.9 with step 0.2. The dashed contour corresponds to $c_m = 0.001$. The concentration maps are shown at the onset of VF (top row), at $t \approx t_{\rm on, rf}$ (middle row), and at $t > t_{\rm on, rf}$.
} 
\label{fig:R-1} 
\end{figure*}

\begin{figure*}
\begin{tabular}{cc}
(a) & (b) \\
\includegraphics[width=0.45\textwidth]{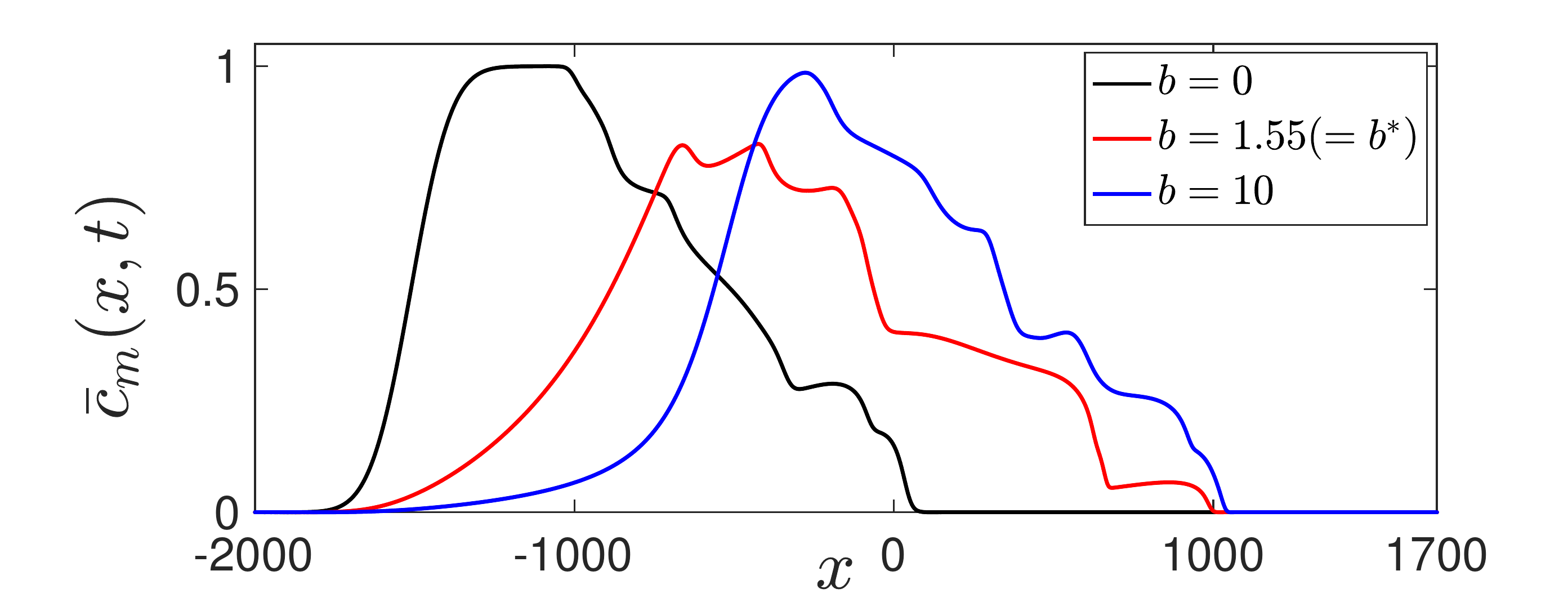} & \includegraphics[width=0.45\textwidth]{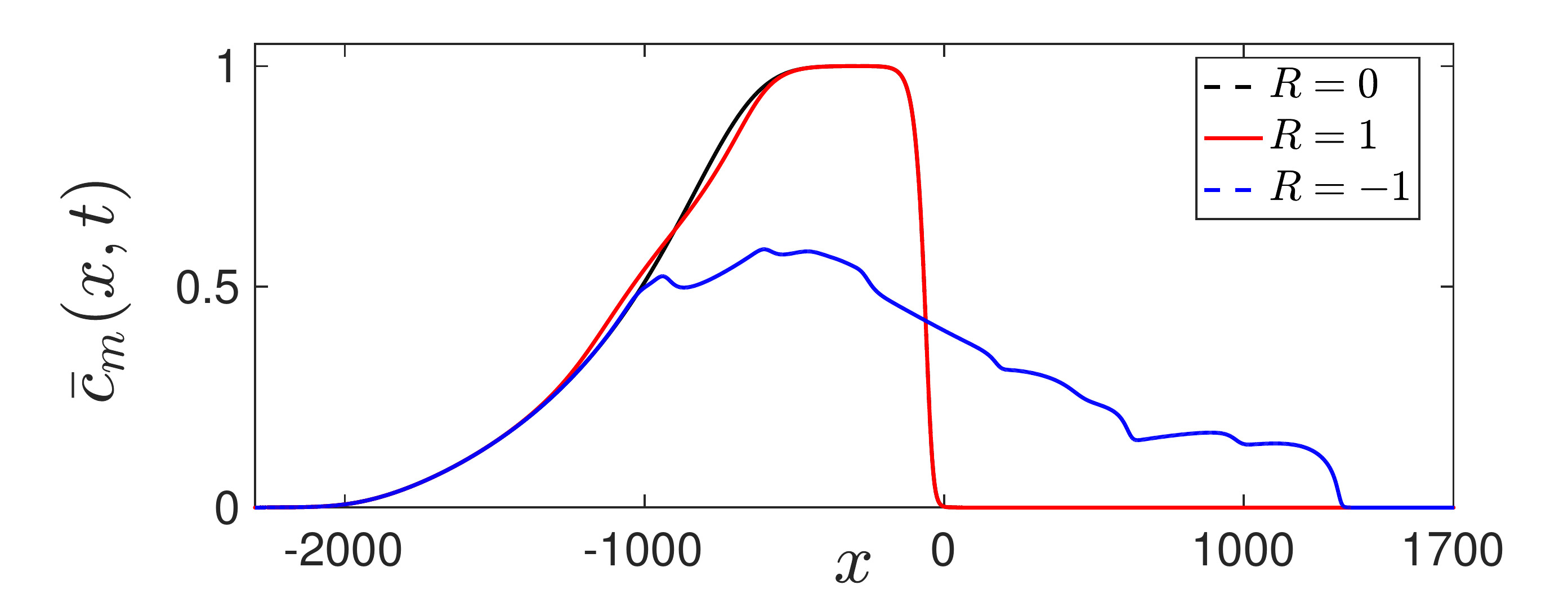}
\end{tabular}
\caption{The transverse averaged concentration profile of the solute $\bar{c}_m(x)$ (a) for different values of $b$ at $t = 6000$ with $k=0.2, R=-1$ (b) for different values of $R$ at $t=8000$ with $k=0.2,b=b^{\star}$.}
\label{fig:avgR-1}
\end{figure*}

\subsection{Interaction between RF and VF for $R<0$}
To analyze the interaction of the RF wave with VF developing at the SL front, we next take $R < 0$ (Figs. \ref{fig:R-1}(a-d)). Fingers originating from the SL front possess both qualitative and quantitative differences compared to the fingers at the RF front. An earlier study explained the self sharpening effect of the SL front with increasing $b$ and its imprint as an early onset of VF \cite{Rana2017a}. To quantify the onset of interaction between VF and RF, we compute $t_{\rm on,rf}$ evaluated the same way as $t_{on,sl}$ and plotted in Fig. \ref{fig:R0_cmx}(b). Interaction of fingers originating from the SL front with the RF front occurs earlier [Figs. \ref{fig:R-1}(b-c)] when compared to the linear adsorption case [Fig. \ref{fig:R-1}(a)]. An early interaction of viscous fingers with the RF front is the resultant of (a) a diminishing separation between the frontal (SL) and rear (RF) interfaces as $b$ increases and (b) an early onset of VF for $b \neq 0$ in comparison to the $b = 0$ case. In combination, these ensure that fingers originating from the SL interface ($b \neq 0$) travel a shorter distance before interacting with the stable rear interface when compared to the linear adsorption isotherm ($b = 0$). This is in strong contrast with the situation of $R > 0$, for which the interaction of the fingered upstream front with the non-fingered downstream front happens at a later time as $b$ increases within a moderate range (see Fig. \ref{fig:R0_cmx}(b)). 

In order to visualize the effect of $b$ on the evolution of
the solute in the case of an unstable frontal interface, the profiles of the transverse averaged concentration $\bar{c}_m(x,t)$ are shown in Fig. \ref{fig:avgR-1}(a) for different values of $b$. Clearly, the concentration profiles for $b\neq 0$ are more distorted than those for $b=0$. Also, for $b=b^{\star}$, the influence of VF on the concentration profiles is more important than for $b=10$. This is due to the fact that $L_{\rm sl}$ is the smallest for $b=b^{\star}$ (see Fig. \ref{fig:slt_vs_b}), which results in a steeper viscosity gradient in comparison to $b=10$ and hence stronger fingering. To compare the influence of positive and negative $R$ on the distribution of the solute concentration, we also plot $\bar{c}_m(x,t)$ in Fig. \ref{fig:avgR-1}(b) for $R=-1,0,1$ at a fixed time $t=8000$ and $b=1.55 ( =b^{\star})$. The fingering instability is more prominent and leads to enhanced spreading for $R=-1$ in comparison to $R=1$ because of the steeper concentration gradient at the shock layer front which favors the VF instability. The other striking difference between $R > 0$ and $R < 0$ cases is the post interaction dynamics of the viscous fingers. As explained earlier, for $R>0$, fingers developing at the rear RF penetrate through the SL, where a steep stable viscosity contrast prevents the breakthrough to occur. The SL is deformed transversely and looses thus its characteristics. In contrast, in a RF front due to Langmuir adsorption ($b \neq 0, k \neq 0$), there is a wider stable zone than for the corresponding linearly adsorbed case ($b = 0, k \neq 0$). This makes it difficult for the backward fingers starting at the SL to reach the upstream end of the stable front [see the overlapping of the concentration profiles at the rear interface in Fig. \ref{fig:avgR-1}(b)]. Thus, a portion of the rarefaction zone preserves the qualitative properties of an expanding wavefront. Moreover, as the stable rear barrier moves closer to the unstable interface with an increasing $b$, backward fingers reorient quicker and spread the solute over a larger zone in the downstream direction [compare Figs. \ref{fig:R-1}(c) and \ref{fig:R-1}(d)].

\begin{figure}
\includegraphics[width=0.45\textwidth]{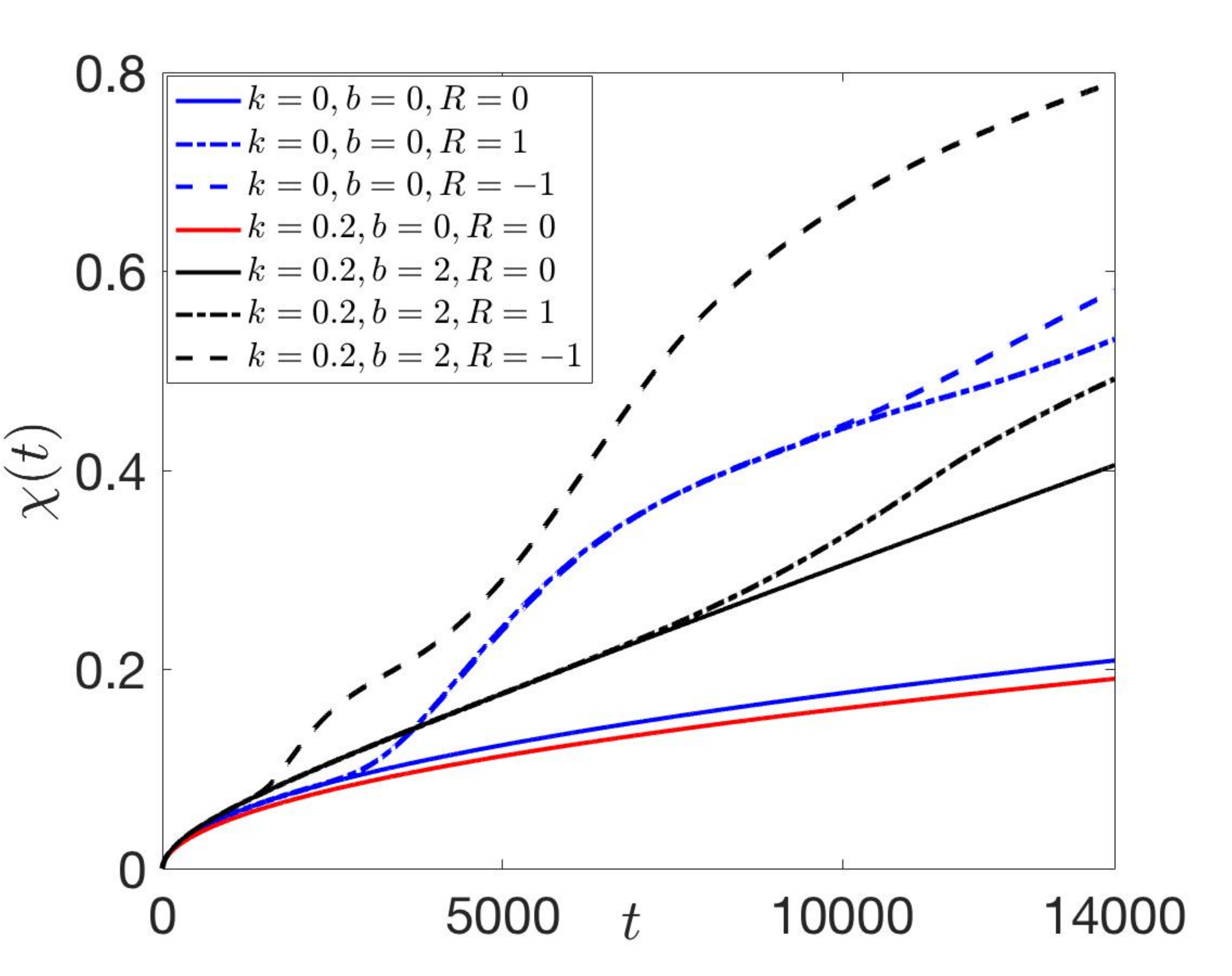} 
\caption{Temporal evolution of the degree of mixing $\chi(t)$ for no-adsorption ($k=0$), linear adsorption ($b=0$) and Langmuir adsorption ($b=2$) cases with $R=0,1,-1$.} 
\label{fig:chi_R0} 
\end{figure}

\begin{figure*}
\centering
(a) \hspace{3.2 in} (b) \\ 
\includegraphics[width=0.45\textwidth]{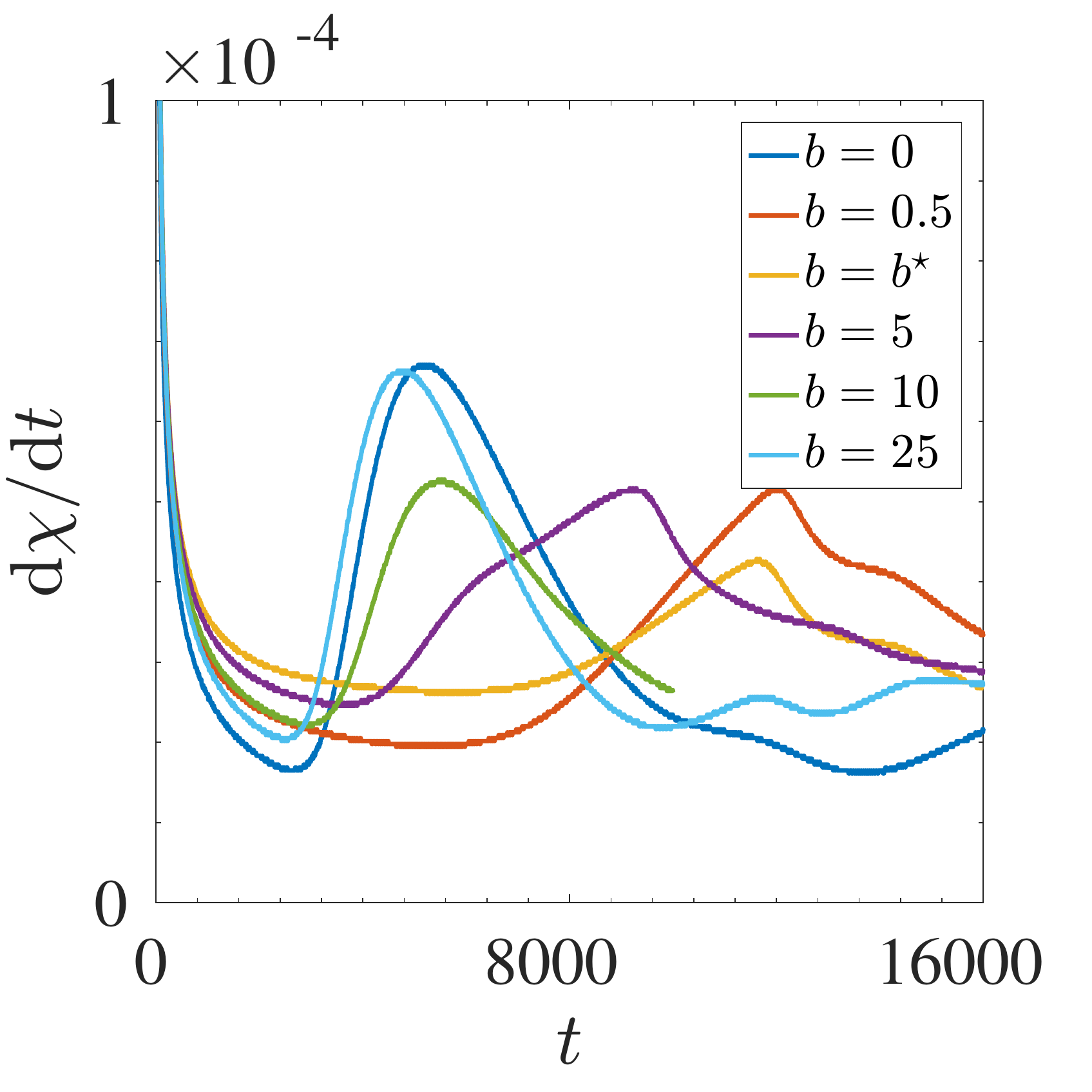}
\includegraphics[width=0.45\textwidth]{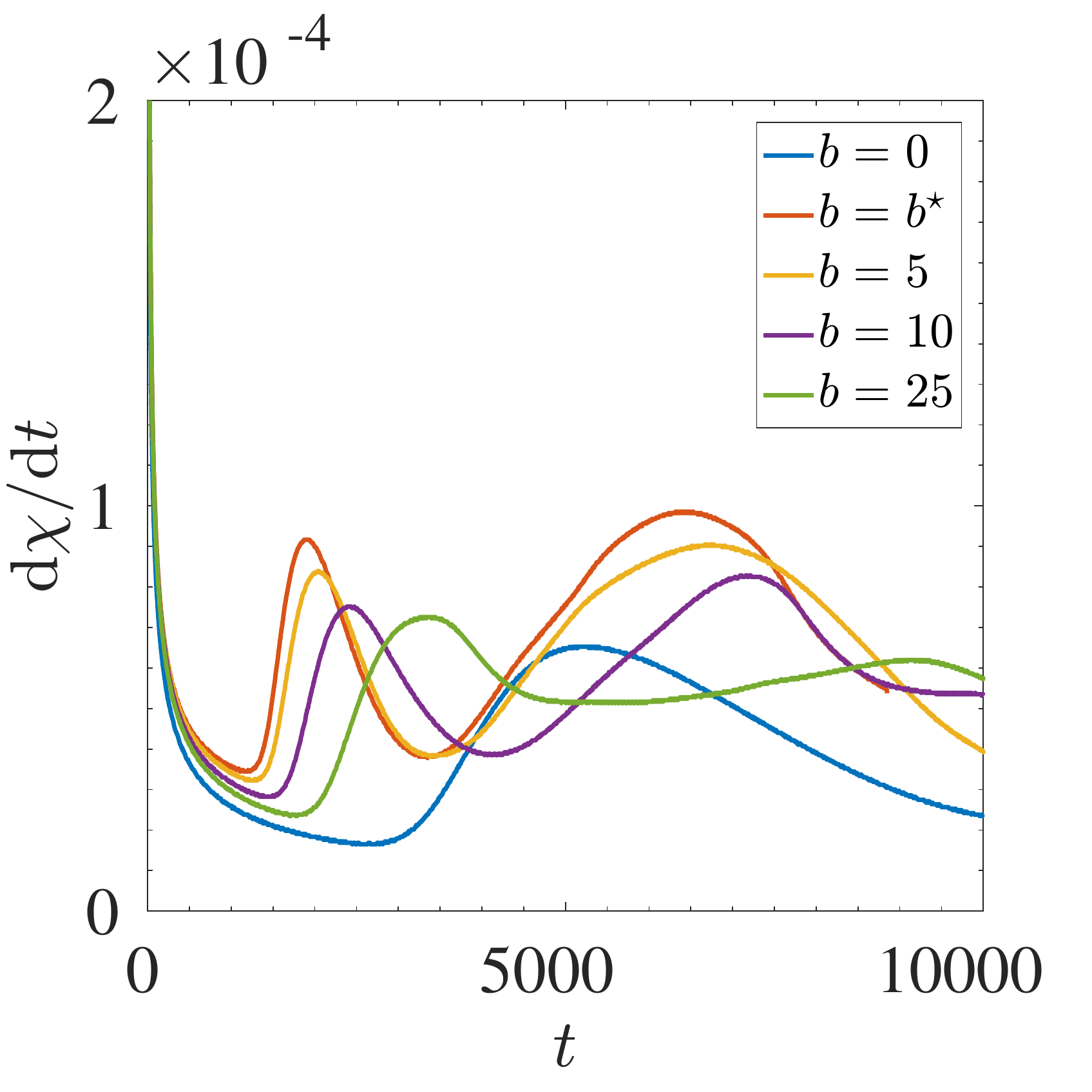}
\caption{(Color online) Temporal evolution of $\dot{\chi}(t)$ for $k = 0.2$ and different values of $b$: $R =$ (a) 1, (b) -1.} 
\label{fig:chi} 
\end{figure*}

\subsection{Degree of mixing \label{sec:deg_mixing}} 
For a more quantitative measure of interaction between nonlinear waves and viscous fingering, we calculate the degree of mixing of $c_m$, defined as $\chi(t) = 1 - \sigma^2(t)/\sigma_{\rm max}^2$, where $\sigma^2(t) = \langle c_m^2 \rangle - \langle c_m \rangle^2$ is the variance of $c_m(x,y,t)$ and $\langle \cdot \rangle$ represents a spatial average \cite{Jha2011}. Here, $\sigma_{\rm max}^2$ corresponds to a perfectly separated state and $\sigma^2=0$ corresponds to a perfectly mixed state which gives $\chi(t)=1$. The degree of mixing $\chi(t)$ is plotted in Fig. \ref{fig:chi_R0} for no-adsorption ($k=0$), linear adsorption ($k \neq 0, b=0$) and Langmuir adsorption ($k \neq 0, b=2$) cases with $R=0,1,-1$. For the stable displacement (i.e. $R=0$), the mixing is enhanced for Langmuir adsorption ($b=2$) in comparison to the no-adsorption ($k=0$) or linear ($b=0$) cases thanks to the formation of the rarefaction zone. In the case with $R=1$ where the rear RF interface is unstable, the rarefaction interface of the Langmuir adsorbed solute has a weaker concentration gradient in comparison to the no-adsorption one. Thus, VF in this RF of the Langmuir adsorbed solute is weaker and hence so is the degree of mixing. However, for the unstable frontal interface i.e. for $R=-1$, the degree of mixing of the Langmuir adsorbed solute increases significantly in comparison to the no-adsorption case. This enhancement in mixing is due to the combined influence of the spreading in the stable rarefaction wave and of the intense fingering due to the sharp viscosity gradient at the unstable frontal interface.

Next, we evaluate the rate of mixing defined as $\dot{\chi} ={\rm d}\chi/{\rm d}t$, plotted in Fig. \ref{fig:chi} for $R=1$ and $-1$. The rate of change, $\dot{\chi}$, is equivalent to the scalar dissipation rate, which corresponds to the homogenization of the fluid due to mixing \cite{Jha2011}. A local minimum (maximum) in the $\dot{\chi}-t$ curve corresponds to an increase (decrease) of the interface length between the solute and the displacing fluid \cite{Jha2011}. For $R = 1, \; k = 0.1$ we plot $\dot{\chi}$ as a function of dimensionless time $t$ for various values of $b$ in Fig. \ref{fig:chi}(a). For $R>0$ and after an initial decrease, the scalar dissipation rate increases after the onset of VF (first local minimum) until the interaction of the fingers with the SL (first local maximum). The wiggles in the $\dot{\chi}-t$ curves at later times are due to splitting and merging of fingers. On the contrary, when $R<0$, two local maxima are observed at $t > 0$ in $\dot{\chi}-t$ curves of a less viscous Langmuir adsorbed sample [Fig. \ref{fig:chi}(b)], the later of which corresponds to $t_{\rm on,rf}$. Similar to a more viscous sample, the first local minimum demarcates the onset of VF. The first maximum is caused by a competition between an enhanced mixing from VF and a reduction in mixing at the expanding RF front. Before the onset of VF, the dissipation rate steeply declines for $b\neq 0$. Nevertheless, we must remember that there can be an opposite contribution from the SL front for the same $b$ values. From these plots, the non-monotonicity of  $t_{\rm on,sl}$ and $t_{\rm on,rf}$ with respect to $b$ are corroborated.

\section{Summary and Conclusions \label{sec:summary}}
Our combined theoretical and computational investigation has focused on the nonlinear dynamics that emerge from the interaction of rarefaction and shock layers and/or viscous fingers that can develop during transport of a finite width sample of solute in a porous medium. 

We have first analyzed the effect of Langmuir adsorption in the absence of VF. As the propagation dynamics of the Langmuir adsorbed solute is influenced by the mobile phase concentration-dependent advection and diffusion of the solute, the semi-infinite solute model admits formation of a shock layer (SL) for a decreasing initial concentration profile \cite{Rana2017a} and a rarefaction (RF) wave for an increasing initial concentration profile \cite{Rana2015}. In the present study we have shown that for finite size samples an interaction between these two nonlinear waves of opposite characteristics occurs after a given time. The non-monotonic variation of the shock layer and rarefaction wave thicknesses with $b$ implies that the shock layer and the rarefaction wave vanish in the limit $b \rightarrow \infty$ and an error function profile then emerges. Our quantification of the interaction between these nonlinear waves further reveals that the interaction onset time has a non-monotonic dependence on the adsorption parameter $b$. Our results show that in order to obtain triangle-like solute profiles the nonlinear adsorption parameters can conveniently be used to tune the interaction of a SL with a RF. 

We have next studied the influence of viscous fingering on the propagation dynamics of finite size samples of non-linearly adsorbed solutes. Our two-dimensional model allows to investigate for the first time the interaction between viscous fingering and nonlinear waves. In the absence of adsorption, the fingering dynamics at the two interfaces corresponding to positive and negative log-mobility ratio are identical until the fingers start interacting with the respective stable interface \cite{Mishra2008}. This is also true in the presence of a linear adsorption isotherm \cite{Mishra2010a}. However, with a Langmuir adsorption isotherm, the fingering dynamics at the frontal interface for a negative log-mobility ratio differs significantly from the fingering dynamics at the rear interface for a positive log-mobility ratio. This is attributed to the fact that the mobile phase solute concentration gradient and hence the viscosity gradient at the frontal interface differs from those at the rear interface.

Our study has also shown significant differences between the interaction of fingers with the SL and the RF wave. When fingers develop on the rear RF front, the steep stable viscosity contrast at the SL front blunts the forward moving fingers that try to intrude the SL which looses its properties due to this interaction. On the contrary, a portion of the rarefaction zone preserves the qualitative property of an expanding wave front when interacting with the VF of the unstable frontal interface developing on the SL. We have shown that, for  Langmuir adsorption ($b \neq 0$), the measurable quantities such as the onset time of interaction of SL with RF, the interaction time of VF with SL, and the interaction time of VF with RF exhibit a non-monotonic variation with respect to $b$ between the two limits of linear adsorption ($b = 0$) \cite{Mishra2007} and the saturated case ($b \rightarrow \infty$) \cite{Mishra2008}. 

The study presented here is of importance in understanding the role of nonlinear adsorption and viscous fingering in various processes where the motion of species results in the formation of nonlinear waves \cite{Abriola,MASLOV2010} or when the motion of species is encountered by nonlinear waves \cite{Ballhaus1974}. Our results could also be helpful in understanding the role of adsorption parameters and viscosity ratios in chromatography applications, CO$_2$ sequestration, subsurface transport as well as in oil recovery techniques. 

\section*{Acknowledgements}CR and AD acknowledge financial support of the FRS-FNRS PDR CONTROL programme. SP acknowledges the support of the Swedish Research Council Grant no. 638-2013-9243. 





%

\end{document}